\documentclass[a4paper,11pt]{article}
\pdfoutput=1

\usepackage{jinstpub}
\usepackage{enumitem}
\usepackage{mathtools}

\newsavebox{\biasTableBox}
\newsavebox{\biasenctable}

\newcommand{\StdL}{StdL}
\newcommand{\StdR}{StdR}
\newcommand{\BC}{BC}
\newcommand{\SGQ}{SGQ}

\title{Sub-electron noise with a buried-channel DePFET: characterisation of novel DePFET concepts}

\author[a,b,1]{Sebastian~M.~Stadler,\note{Corresponding author.}}
\author[b]{Alexander~B\"ahr,}
\author[b]{Jelena~Ninkovi\'c,}
\author[b]{Peter~Lechner,}
\author[b]{Johannes~Treis}
\author[b]{and Christian~Sandow}

\affiliation[a]{Munich University of Applied Sciences, Center for Physical and Biomedical Engineering (CPaB),\\
	Lothstra\ss e 34, 80335 Munich, Germany}
\affiliation[b]{Semiconductor Laboratory of the Max Planck Society (HLL),\\
	Isarauenweg 1, 85748 Garching, Germany}

\emailAdd{ses@hll.mpg.de}

\abstract{Depleted P-channel Field Effect Transistor (DePFET) pixel sensors combine Fano-limited energy resolution with fast, low-noise readout and are used in instruments ranging from particle tracking to X-ray astronomy. In this work, two novel DePFET technologies that address the two factors of the signal-to-noise ratio through fundamental changes to the intrinsic design are characterised: a buried-channel design aimed at reducing the limiting low-frequency noise by weakening the coupling between the channel and Si/SiO$_2$ interface traps, and a super-$g_q$ design intended to increase the charge gain by reducing the effective internal-gate capacitance. Their noise and energy resolution were mapped across the transistor bias space, over a wide temperature range ($100$--$260\,\mathrm{K}$) and for integration times of $0.4$--$9\,\upmu\mathrm{s}$ (readout times of $2$--$19.2\,\upmu\mathrm{s}$ per row), and compared with standard DePFETs as developed for the NewAthena Wide Field Imager (WFI). All three variants reach their lowest noise at about $155\,\mathrm{K}$. The buried-channel device achieves an excellent equivalent noise charge (ENC) of $0.96\,\mathrm{e}^-$ (at most $0.83\,\mathrm{e}^-$ after subtracting the readout-electronics noise), demonstrating sub-electron noise over a complete $64\times64$ DePFET matrix in a single readout, alongside a power consumption reduction of more than $50\%$. The super-$g_q$ DePFET shows a $2.6$-fold increase in charge gain (about $1100\,\mathrm{pA/e^-}$) but exhibits higher total noise due to an excess noise contribution. The results highlight the potential of the novel DePFETs, with the buried-channel technology offering immediate benefits for low-noise applications such as the WFI.}


\begin{document}
	\maketitle
	\flushbottom
	
	\section{Introduction}
	\label{sec:introduction}
	Driven by the aspiration to answer new scientific questions, the requirements placed on silicon detectors for spectroscopy and imaging continue to grow. In this context, low noise is essential for precise charge measurement, since it sets the smallest signal that can be reliably distinguished from the noise floor and thus the achievable signal-to-noise ratio (SNR). In spectroscopy, low noise sharpens the energy resolution of spectral lines and lowers the minimum detectable energy, whereas in imaging it preserves faint sources and low-contrast features that would otherwise be lost. Beyond these direct benefits, a low noise floor also relaxes system-level design trade-offs. Smaller pixel pitches become feasible despite increased charge sharing, whereas shorter readout times support higher frame rates, mitigate pile-up, and decrease the accumulation of leakage charge, thereby facilitating operation at elevated temperatures. Applications that require thin sensors or have a limited material budget, and therefore exhibit lower charge deposition, also benefit.
	
	It is within this tension between competing detector parameters that further development of the DePFET (Depleted P-channel Field Effect Transistor) sensor becomes especially valuable. Owing to their adaptability to high frame rates, high spatial resolution, Fano-limited energy resolution, and wide dynamic range, DePFETs have been selected for a broad range of current and planned instruments~\cite{andricekDePFETRecentDevelopments2022}. These are the PXD in the Belle II experiment for particle tracking~\cite{ahlburgNewCompleteBelle2024}, the DePFET-based MIXS sensor onboard the planetary science mission BepiColombo for X-ray spectroscopy~\cite{treisMIXSBepiColomboIts2010}, the development of ultrafast real-time imagers for direct electron detection (EDET) for transmission electron microscopes (TEMs)~\cite{predikakaEDETDH80kCharacterization2020} or the development of the Wide Field Imager (WFI) of the next generation X-ray imaging spectroscopy onboard the NewAthena satellite~\cite{treberspurgCharacterization2562562020}.
	
	Building on this diverse foundation, a recent R\&D initiative at the Semiconductor Laboratory of the Max Planck Society (HLL) explores the untapped potential of DePFET design considerations. The production comprises variations of the standard DePFETs based on the current NewAthena layout, featuring optimised implantation parameters as well as additional implants such as HALO and deep n-type implants for sharper pixel boundaries, with pixel pitches ranging from $15\,\upmu\mathrm{m}$ to $130\,\upmu\mathrm{m}$, but also novel DePFET concepts were realised. Alongside implementations in various gate dimensions, optimised Infinipix and Quadropix layouts, RNDR devices with an improved transfer channel, on-pixel frame store structures, and MOS-CCD test devices with DePFETs at the readout node were fabricated.
	
	\begin{figure}[htbp]
		\centering
		\raisebox{0.14\height}{
			\includegraphics[width=.4\textwidth,trim=60 70 50 70,clip]{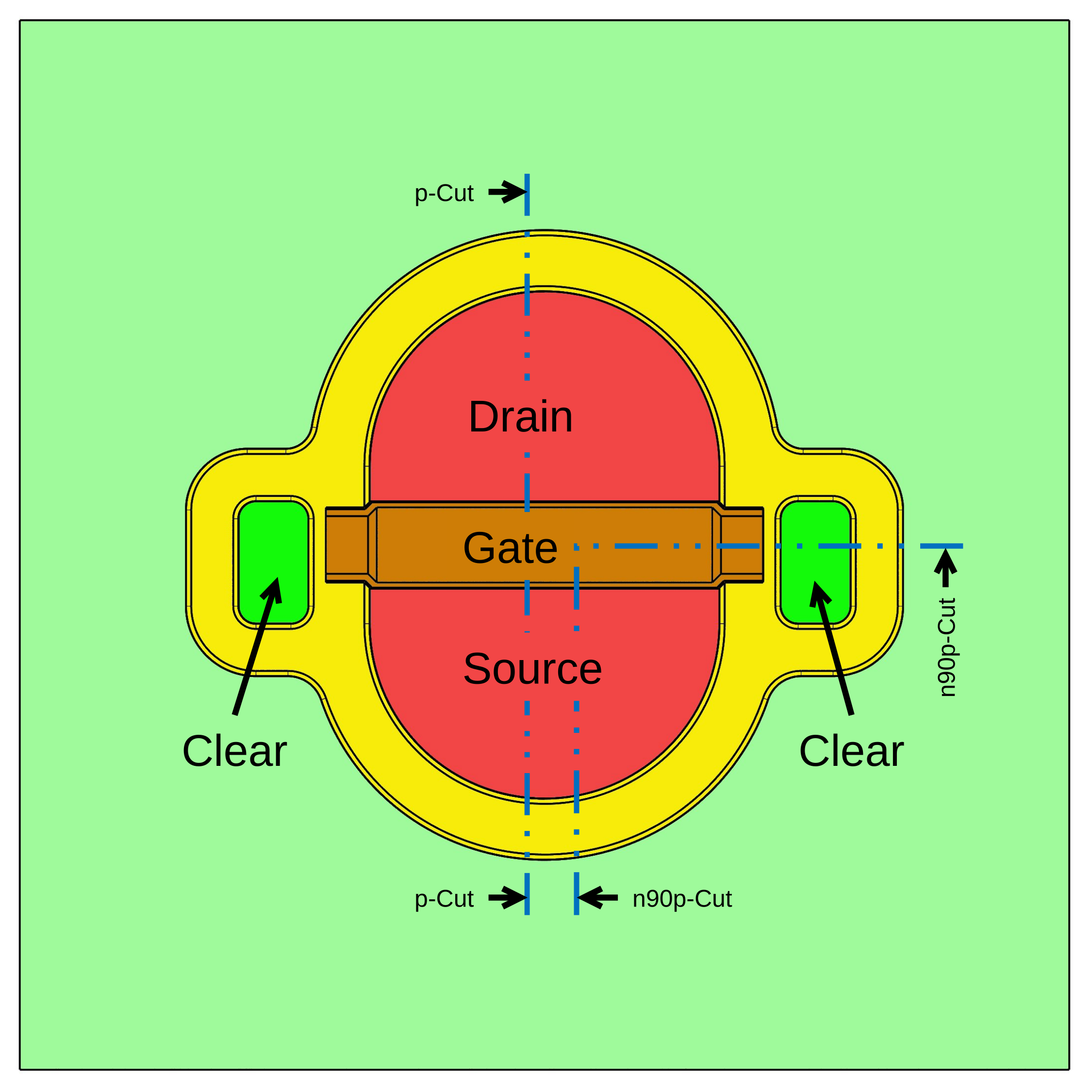}
			
		}
		\qquad
		\includegraphics[width=.5\textwidth]{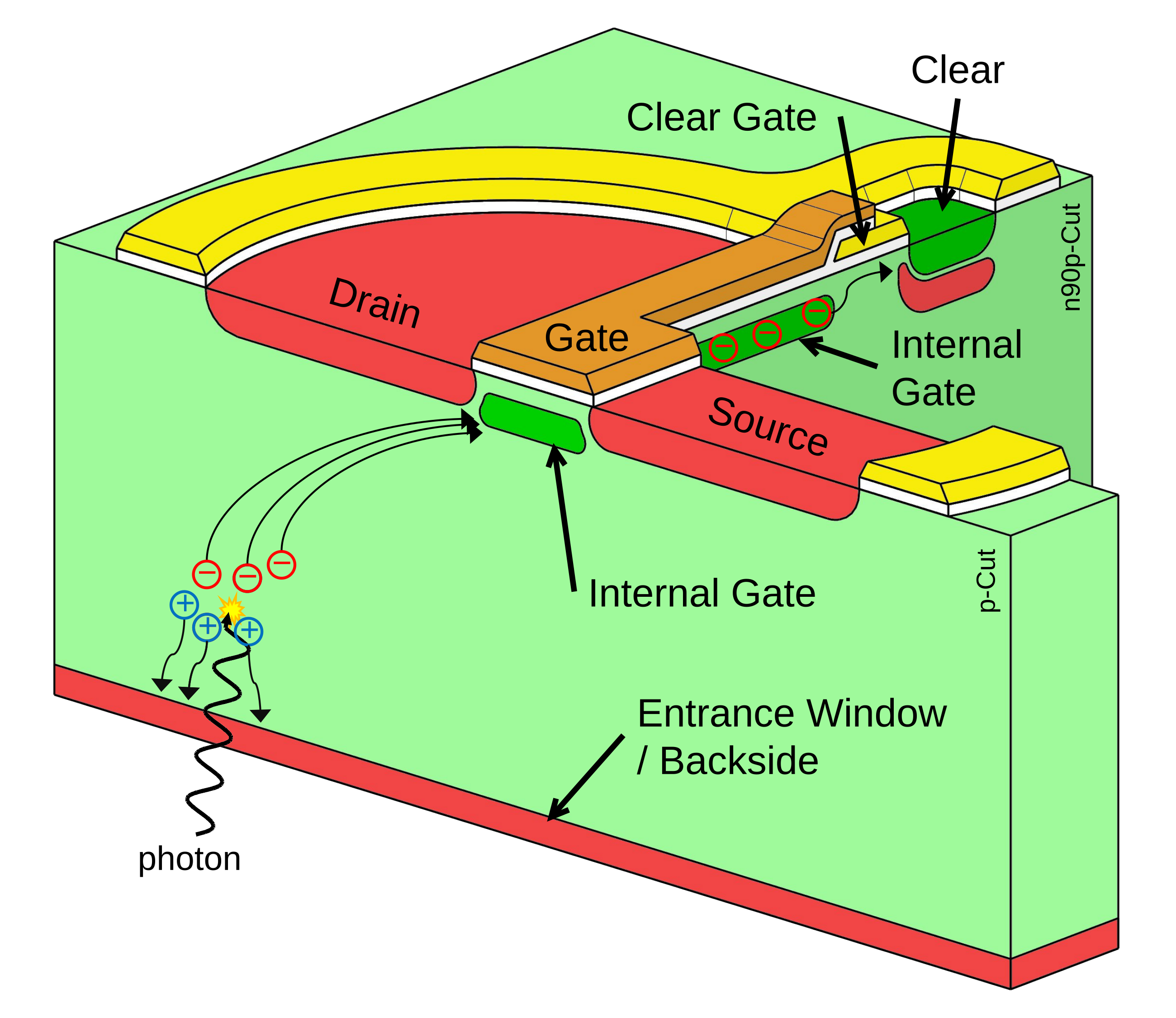}
		\caption{\label{fig:depfetCut} Schematic drawing of the standard DePFET layout in top view (left) and in 3D (right), with cuts along the p-channel (p-Cut) and through the clear-structure (n90p-Cut). The p-channel cut illustrates electron-hole pair generation by an X-ray interaction, with arrows indicating the drift of electrons to the internal gate and of holes to the backside, whereas the clearing of the electrons accumulated in the internal gate is depicted in the clear-structure cut. The Clear is embedded in a deep p-well (red), which forms a potential barrier towards the bulk.}
	\end{figure}
	
	The basic layout of a standard DePFET technology is shown in figure~\ref{fig:depfetCut} in top view (left) and in 3D with cuts along the p-channel and through the clear-structure (right). It consists of a p-channel MOSFET on a high-resistivity, n-doped silicon substrate. A shallow channel implant adjusts the threshold voltage and improves its pixel-to-pixel uniformity, while a deep n-implant beneath the transistor channel forms a potential minimum for electrons. By applying a sufficiently high reverse bias to a thin p+ doped layer on the backside, which also serves as the entrance window, the substrate is completely depleted by means of sidewards depletion, rendering the entire volume sensitive. Absorbed ionising radiation generates electron-hole pairs that separate in the depletion field: holes drift to the back contact, while electrons drift to the potential minimum, where they accumulate. The collected electrons induce mirror charges in the p-channel, modulating its conductivity. In analogy to an external gate voltage, the deep n-implant is commonly referred to as the internal gate (IG), because of this steering effect on the channel current. The intrinsic, or internal, amplification is quantified by the charge gain, defined as the change in drain current per collected electron and is denoted as $g_q$. The internal gate can be reset on demand by applying adequate voltage pulses to the adjacent Clear and Clear-Gate contacts. This creates a potential gradient to the Clear that drains the accumulated electrons.
	
	The presented work investigates the noise and spectral performance of two novel DePFET technologies, the buried-channel DePFET and the super-$g_q$ DePFET, each addressing one of the two factors of the SNR, and compares them with the standard technology. Their design differences are described in section~\ref{sec:designs} and the measurement system and the analysis in section~\ref{subsec:setup_readout}. A bias configuration for the lowest noise is established in section~\ref{sec:opti}, covering the depletion and clear-process voltages and the optimum intrinsic DePFET transistor settings. Their noise behaviour is further explored with respect to temperature and integration time (sections~\ref{sec:temperature} and~\ref{sec:optimal}). Common measurement limitations are discussed in section~\ref{sec:discussion}, together with the advantages and drawbacks of the novel technologies. The main findings are summarised in section~\ref{sec:summary}. 

	\section{Novel DePFET designs}
	\label{sec:designs}
	The investigated novel DePFET technologies differ from the standard DePFET technology only in their intrinsic design.
	\paragraph{Super-$g_q$ DePFET.} Due to the DePFET's low-noise internal amplification, increasing $g_q$ is one way to improve the SNR~\cite{andricekDePFETRecentDevelopments2022}. In a first-order expression, the charge gain is set by the transconductance $g_\mathrm{m}$ of the DePFET transistor and by the coupling between the internal gate and the channel,
	\begin{equation}
		\label{eq:gq_gm}
		g_q = \frac{q\,f}{C_\mathrm{IG}}\,g_\mathrm{m},
	\end{equation}
	where $q$ is the elementary charge, $C_\mathrm{IG}$ is the effective internal-gate capacitance and $f$ ($0 < f \leq 1$) a device-specific factor that quantifies the coupling efficiency from IG to the channel~\cite{lutzSemiconductorRadiationDetectors2008}. In the ideal mirror-charge case, $f \to 1$. In practice $f < 1$, due to non-ideal effects such as stray capacitances. In the following, $q\,f/C_\mathrm{IG}$, i.e. $g_q/g_\mathrm{m}$, is referred to as the internal-gate coupling. The $g_q$ in saturation can be obtained by inserting the transconductance of the ideal MOSFET model,
	\begin{equation}
		\label{eq:gq_eq1}
		g_q = \frac{q\,f}{C_\mathrm{IG}}\sqrt{2\mu C_\mathrm{ox}'\frac{W}{L}I_\mathrm{DS}},
	\end{equation}
	where $\mu$ is the carrier mobility, $C_\mathrm{ox}'$ the oxide capacitance per unit area, $W$ and $L$ channel width and length and $I_\mathrm{DS}$ the drain current. Since the mirror charge induced in the channel acts on the drain current like a charge induced by the external gate, $C_\mathrm{IG}$ is given by the gate capacitance over the area of the internal gate. In standard DePFET technology, the IG length is determined by the spacing between the source/drain implants, which in turn are self-aligned to the external gate edges. The internal gate width is constrained to the external gate width through the adjacent Clear-Gate structure (figure~\ref{fig:depfetCut}). Therefore, with the approximation $C_\mathrm{IG} \approx C_\mathrm{ox}'WL$, eq.~\eqref{eq:gq_eq1} becomes
	\begin{equation}
		\label{eq:gq_eq2}
		g_q \approx q\,f \sqrt{\frac{2\mu I_\mathrm{DS}}{C_\mathrm{ox}' W L^3}}.
	\end{equation}
	Accordingly, $g_q$ depends most strongly on the gate length, scaling as $L^{-3/2}$, whereas the width enters only through $W^{-1/2}$. Previous studies have shown that $g_q$ can be improved by shortening the gate and therefore the internal gate, thereby reducing its effective capacitance~\cite{rummelIntrinsicPropertiesDEPFET2009}. A considerable reduction of $L$, however, increases the lateral electric field at the drain end of the channel, which gives rise to hot-carrier effects such as impact ionisation. As a result, the noise performance deteriorates and the SNR improvement from larger $g_q$ is partially offset. Mitigating high-field effects generally requires a lower $\lvert V_\mathrm{DS} \rvert$. Maintaining saturation then restricts the available overdrive according to $\lvert V_\mathrm{DS} \rvert \gtrsim \lvert V_\mathrm{ov} \rvert$, thereby narrowing the usable bias range. Moreover, the internal gate couples increasingly to the source and drain regions due to the parasitic overlap and fringing capacitances, so that $f \downarrow$. Consequently, the benefit of further scaling falls below the ideal $g_q \propto L^{-3/2}$ assumption.
		
	\begin{figure}[htbp]
		\centering
		\includegraphics[width=.29\textwidth,clip]{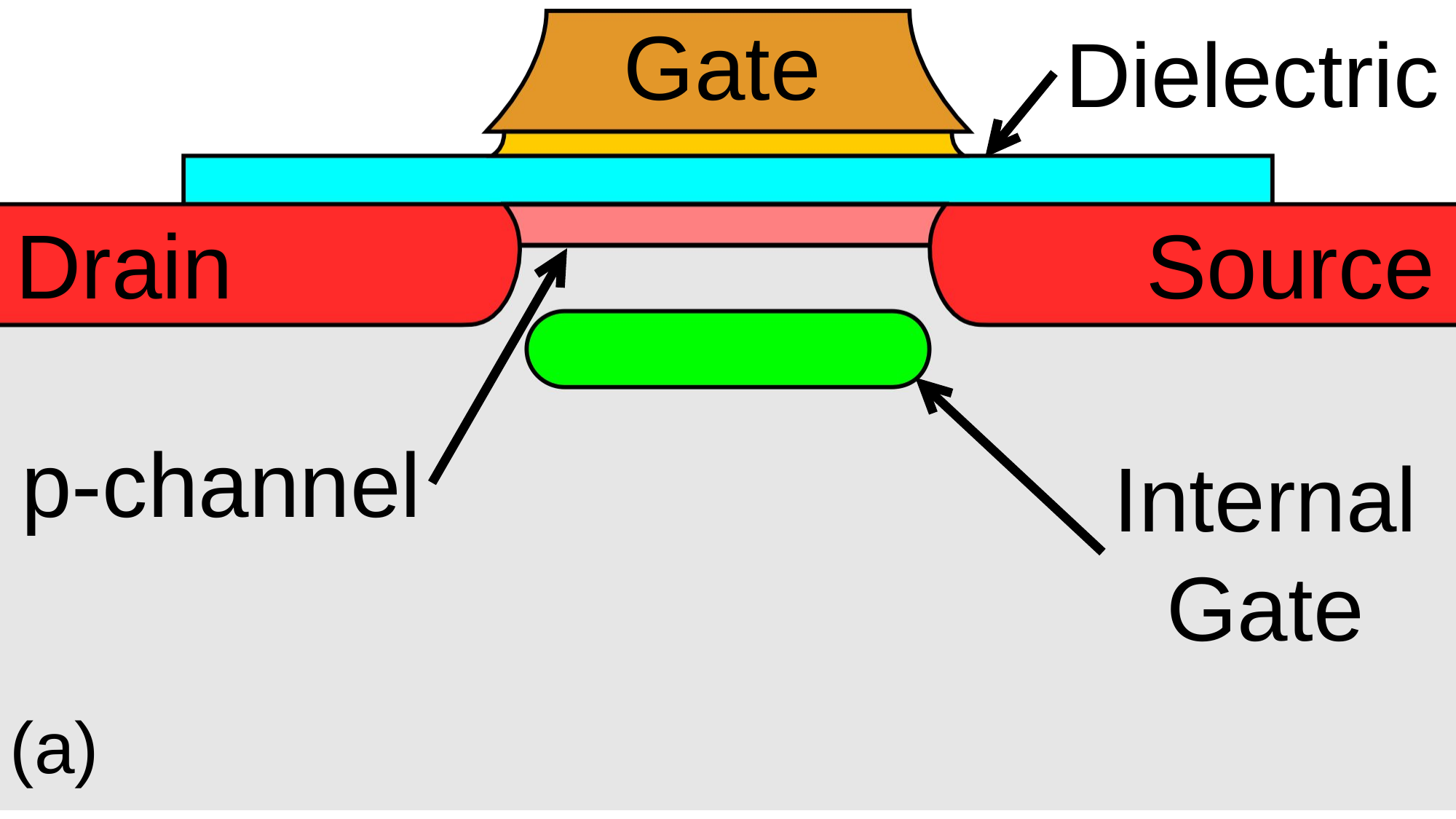}
		\qquad
		\includegraphics[width=.29\textwidth,clip]{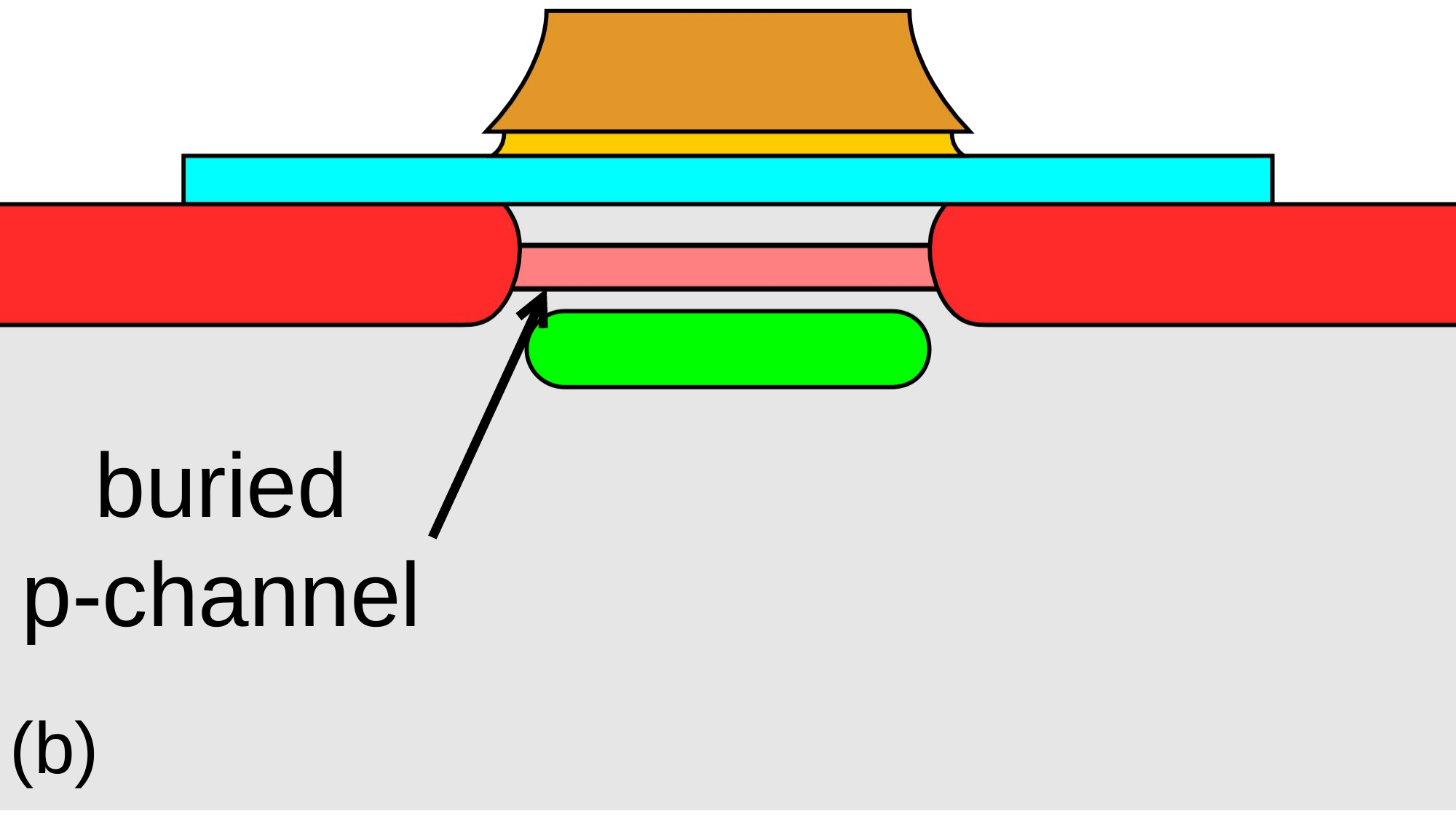}
		\qquad
		\includegraphics[width=.29\textwidth,clip]{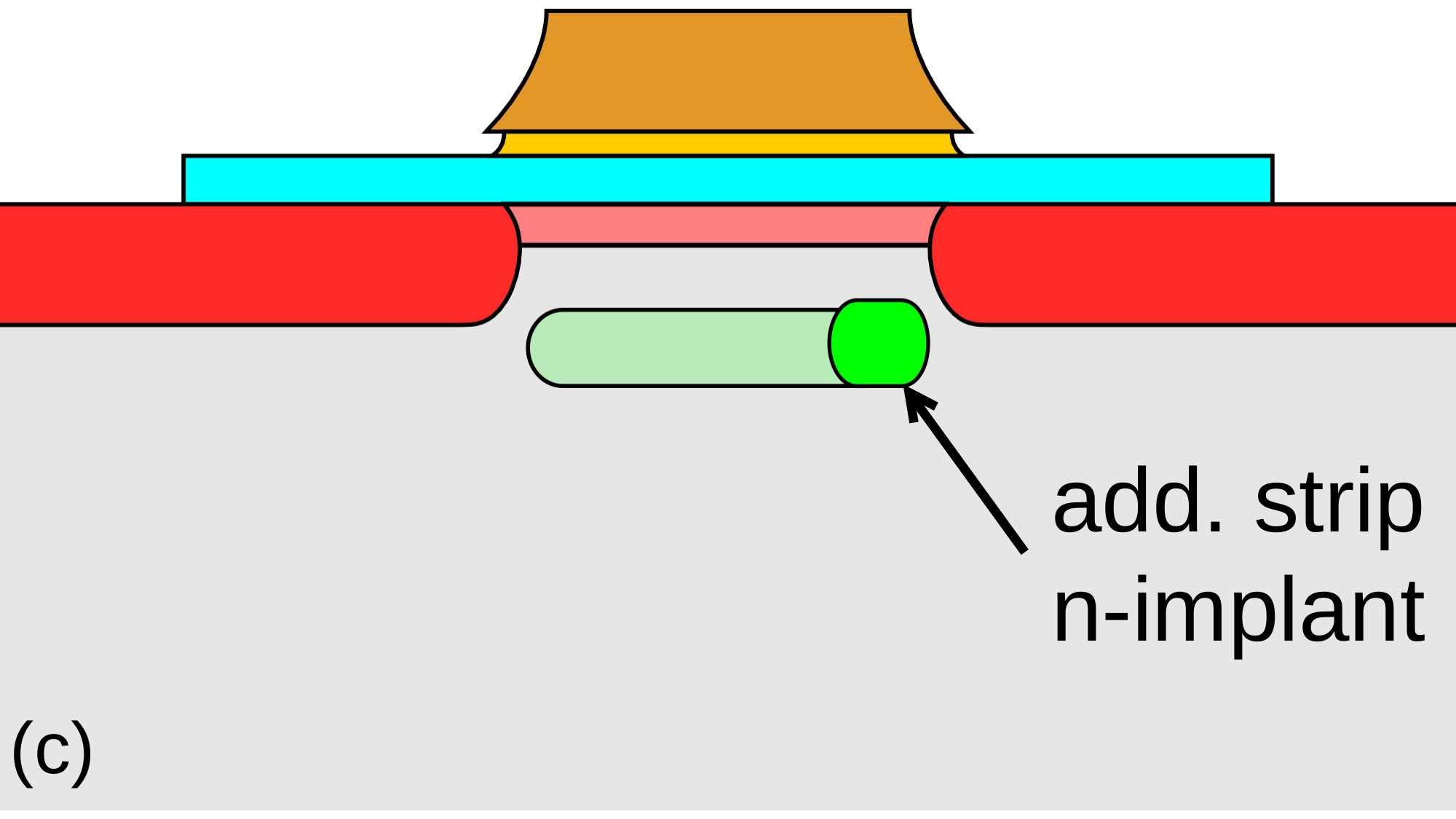}
		\caption{\label{fig:depfetDesigns} Schematic cross-sections of the intrinsic design of the three DePFET variants used in this work: (a) standard DePFET, with a shallow p-implantation at the Si/SiO$_2$ interface and the internal gate size defined by the source-drain spacing; (b) buried-channel DePFET, with the p-implantation shifted further away from the Si/SiO$_2$ interface deeper into the bulk; (c) super-$g_q$ DePFET, with an additional strip-shaped n-implant on the source side, which appears as a localised dot in the 2D cross-section.}
	\end{figure}
	
	The super-$g_q$ DePFET technology addresses this limitation by introducing an additional strip-shaped n-type implant on the source side (figure~\ref{fig:depfetDesigns} (c))~\cite{bahrAdvancedDePFETConcepts2020}. This confines the collected charge to a narrow region beneath the channel and thereby reduces the internal-gate length and thus its effective IG capacitance, while the external-gate length remains unchanged, effectively making the internal-gate length independent of the external gate. Since the implant is localised on the source side and the deep n-implant is reduced elsewhere, the potential distribution along the channel changes. When the internal gate is cleared, the positive space charge lowers the local channel hole density, causing the channel resistance, and with it the potential drop across the source-side section, to increase. Furthermore, the strip implant is located closer to the channel. Its space charge thus exerts a stronger field on the channel and also compensates to a small extent the shallow p-channel doping, which was therefore chosen slightly higher. According to TCAD device simulations, these super-$g_q$ modifications improve the charge gain threefold compared to the standard technology~\cite{stadlerFirstInsightsSupergq2024b, bahrAdvancedDePFETConcepts2020}.

	\paragraph{Buried-channel DePFET.} The minimum attainable noise level of a single readout is dominated by low-frequency ($1/f^{\alpha}$) noise.\footnote{Assuming sufficient cooling, so that bulk-leakage noise is negligible.} Subsequent signal processing can suppress this contribution only to a limited extent~\cite{radekaSignalProcessingParticle2020}. It is therefore addressed at its source by optimising the intrinsic structure of the DePFET. Besides improving the Si/SiO$_2$ interface, an alternative approach is to shift the conductive channel away from the interface and thereby weaken the electrostatic coupling between charge carriers in the channel and traps at the interface~\cite{wilcoxLowtemperatureCharacterizationBuriedchannel1989}. A DePFET with a buried channel is realised simply by repositioning the shallow p-implant deeper below the gate (figure~\ref{fig:depfetDesigns} (b)). When operated in buried-channel mode (i.e. no inversion layer is formed at the Si/SiO$_2$ interface), noise comparable to that of a JFET is reachable~\cite{kandiahImprovementsLowFrequency1990}. In this operating mode, the depleted silicon layer between the interface and the channel contributes a capacitance in series with the gate dielectric~\cite{huDesignTradeoffsSurface1985b}. Consequently, $C_\mathrm{ox}'$ must be replaced in eqs.~\eqref{eq:gq_eq1} and~\eqref{eq:gq_eq2} by the gate-channel capacitance,
	\begin{equation}
		\label{eq:CG}
		\frac{1}{C_\mathrm{G}'} = \frac{1}{C_\mathrm{ox}'} + \frac{x_\mathrm{s}}{\varepsilon_\mathrm{Si}},
	\end{equation}
	where $\varepsilon_\mathrm{Si}$ is the permittivity of silicon and $x_\mathrm{s}$ the effective thickness of the depleted layer, which depends on $V_\mathrm{GS,on}$. For $x_\mathrm{s} \to 0$, the standard case $C_\mathrm{G}' = C_\mathrm{ox}'$ is recovered. In buried-channel mode, $C_\mathrm{IG} \approx C_\mathrm{G}'WL$ is therefore smaller and depends on $V_\mathrm{GS,on}$.
		
	\subsection{Devices under test (DUT)}
	\label{subsec:DUT}	
	Four devices were characterised: a buried-channel device (\BC{}) and a super-$g_q$ device (\SGQ{}), as well as two devices based on standard technology, a reference device (\StdR{}) from the same R\&D production and a legacy device (\StdL{}) from the latest WFI production. \StdR{} provides a direct comparison within the R\&D production, eliminating inter-production variations, while \StdL{} has been extensively characterised in previous campaigns and was also used to validate the measurement setup~\cite{treberspurgCharacterization2562562020, bonholzerDrainCurrentCharacteristics2022, sommerCharacterisationNewAthenaWFIs2026b}. All pixel matrices presented in this work share the same physical layout, with a design gate length and width of $5\,\upmu\mathrm{m}$ and $20\,\upmu\mathrm{m}$, respectively, and a pixel pitch of $130\,\upmu\mathrm{m}$. The matrices in $64 \times 64$ format were fabricated at HLL on $450\,\upmu\mathrm{m}$ thick wafers. In addition, each sensor chip houses a temperature diode. 
	
	\section{Measurement system and readout chain}
	\label{subsec:setup_readout}
	A dedicated measurement system was developed to characterise the DePFET sensors over a wide temperature range and under representative operating conditions. The detector assembly is a replaceable component of the measurement setup, placed inside a vacuum chamber. It consists of the DePFET sensor, which is glued onto a ceramic carrier and wire-bonded to a printed circuit board (PCB), as well as to the Switcher(A) and the Veritas(2.1) readout ASIC, which was developed for the WFI~\cite{herrmannVERITAS22Low2018a}, mounted on that board. Plastic standoffs secure the PCB to the carrier and ensure sufficient thermal insulation. In addition, resistors provide active heating, allowing the electronics to be maintained at a stable, elevated temperature while the carrier with the sensor is cooled by a cryocooler. A radioactive source for calibration and spectral performance measurements can be moved in front of the entrance window. 
	
	\begin{figure}[htbp]
		\centering
		\includegraphics[width=.60\textwidth,clip]{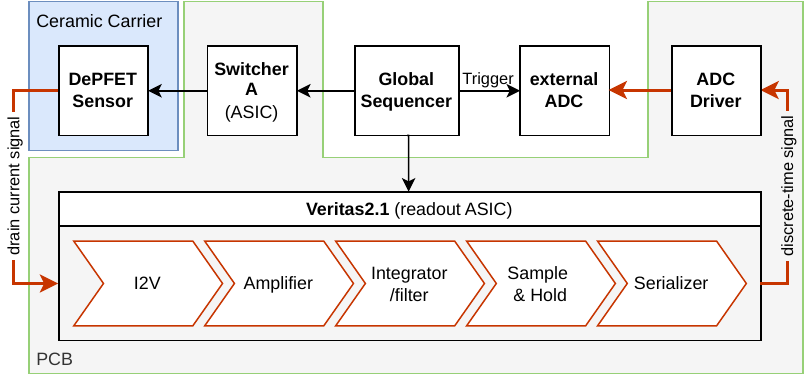}
		\hspace{.02\textwidth}
		\includegraphics[width=.36\textwidth,trim=0 0 12 0,clip]{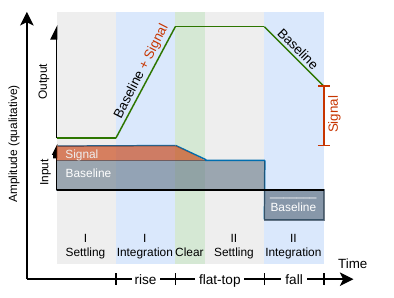}
		\caption{\label{fig:Readout} Schematic illustration of the measurement system. Left: detector assembly and its components. Black arrows indicate control and orange arrows the signal path. Right: dual-slope integration (DSI) in the Veritas integrator/filter stage. After the row is switched on, an initial settling period (I) of $0.6\,\upmu\mathrm{s}$ is required. Then the baseline together with the DePFET signal is integrated (rise), after which at the flat-top a clear pulse removes charge from the IG, followed by a settling period (II). The second integration (fall) is equal in duration and samples the baseline alone but with inverted polarity, thereby subtracting it from the first integration. The output signal is proportional to the signal current.}
	\end{figure}
	
	The DePFET pixel matrix is operated in drain readout mode within a rolling-shutter scheme, where one row is addressed after another by the Switcher. All pixels in the active row are read out in parallel through the 64 channels of the Veritas, while the matrix remains sensitive. In each channel, the DC component of the drain current is removed by a constant subtraction current, identically applied to all channels, before the residual current is converted by a transimpedance amplifier stage. The resulting voltage signal is amplified, and then passed on to the integrator/filter stage, where the pixel-specific baseline subtraction is performed by dual-slope integration (DSI), which acts as a time-variant filter on the conditioned signal with a trapezoidal weighting function. The first integration samples the baseline together with the signal contribution, after which a clear pulse is applied during the flat-top to remove the collected charge from the internal gate. The second integration samples the baseline alone in inverted polarity. The remaining signal is therefore proportional to the signal current (figure~\ref{fig:Readout} right). The result is stored in the sample-and-hold stage, before the complete row is serialised, buffered by an ADC driver (ADA4932-2), and transmitted to the 16-bit external ADC (M2i.4961), concluding the readout chain. The timing parameters are controlled by a global sequencer, which also initiates the clearing process via the Switcher (figure~\ref{fig:Readout} left).
	
	\subsection{Measurement settings and analysis}
	\label{subsec:settings_analysis}
	The I2V (current-to-voltage), preamplifier, and integrator/filter stages of the Veritas each offer several user-selectable gain settings. For all presented measurements, the lowest gain configuration was used for better comparability and to ensure sufficient headroom in terms of dynamic range, unless otherwise stated.\footnote{$\mathrm{I2V} = 0\mathrm{b}00$ (table~\ref{tab:I2V}), preamplifier gain $= 2\times$.} 
	
	A single measurement comprised a series of dark frames and frames illuminated by a $^{55}\mathrm{Fe}$ calibration source.\footnote{The number of frames depended on several factors. It was chosen such that the statistical errors arising from the finite sample size were below $0.05\%$ of $\mathrm{FWHM_{all}}$, $0.15\%$ of $\mathrm{FWHM_{sng}}$, and $0.006\%$ of the median ENC.} The acquired frames were analysed with the ROAn framework (ROOT-based Offline and Online Analysis, version~0.8/01)~\cite{laufROOTBasedOffline2014}. Both dark and illuminated frames were corrected for offset and common-mode variations. The noise level of each pixel was then obtained from the standard deviation of the corrected dark frames. Using the Mn~K$\alpha$ peak in the spectrum of the illuminated frames, the energy calibration gain $G_\mathrm{E}\,(\mathrm{in ADU/eV})$ was determined for each pixel.\footnote{The doublet-weighted Mn~K$\alpha$ energy of $5895\,\mathrm{eV}$ was used.} Together with the temperature-dependent pair generation energy $w(T)$~\cite{loweMeasurementElectronHole2007b}, the output-referred noise ($\mathrm{Noise_{out}}$) was converted from analogue-to-digital unit (ADU) to the input-referred noise ($\mathrm{Noise_{in}}$), expressed as the total equivalent noise charge ($\mathrm{ENC_{tot}}$, in short ENC) and reported as the median over all valid pixels. For a fixed signal, the Mn~K$\alpha$ photon energy ($E_{\mathrm{K}\alpha}$), the output signal in ADU changes only with $G_\mathrm{E}$, which equally scales $\mathrm{Noise_{out}}$. SNR is therefore inversely proportional to ENC,
	\begin{equation}
		\label{eq:snr}
		\mathrm{SNR}_{\mathrm{K}\alpha} = \frac{G_\mathrm{E}\,E_{\mathrm{K}\alpha}}{\mathrm{Noise_{out}}} = \frac{G_\mathrm{E}\,E_{\mathrm{K}\alpha}}{G_\mathrm{E}\,w(T)\,\mathrm{ENC}} = \frac{E_{\mathrm{K}\alpha}}{w(T)\,\mathrm{ENC}}.
	\end{equation}
	$G_\mathrm{E}$ relates to $g_q$ by	
	\begin{equation}
		\label{eq:GE_to_gq}
		g_q = \frac{w(T)\,G_\mathrm{E}}{A},
	\end{equation}
	where $A$ (in ADU/pA) is the current-to-ADU conversion of the readout chain for the selected Veritas gain settings and integration time.\footnote{The conversion factor $A$ was evaluated using the Test-IN of the Veritas on a dedicated detector assembly without a DePFET sensor populated.} Dividing $\mathrm{Noise_{out}}$ by $A$ yields the noise referred to the sensor output ($\sigma_\mathrm{out}$), independent of the Veritas gain settings. From eqs.~\eqref{eq:snr} and~\eqref{eq:GE_to_gq}, it follows that
	\begin{equation}
		\label{eq:enc_sigma_gq}
		\mathrm{ENC} = \frac{\sigma_\mathrm{out}}{g_q}.
	\end{equation}
	
	The energy resolution, expressed as the full width at half maximum (FWHM) of a Gaussian fit to the Mn~K$\alpha$ doublet, was determined either over all valid events, including multi-pixel clusters due to charge sharing ($\mathrm{FWHM_{all}}$), or solely over single-pixel events ($\mathrm{FWHM_{sng}}$). A maximum signal charge spread over $2 \times 2$ pixels was assumed, considering the pixel size and sensor thickness. Events with larger signal patterns were discarded. 
	
	Bad pixels, flagged by ROAn due to an abnormal offset, noise or signal response, or due to an insufficient clear, were excluded from the analysis. Remaining pixels were classified as valid. Measurements in which the number of bad pixels exceeded the device-specific baseline by more than 200 were discarded entirely to ensure a comparable population of valid pixels across all measurements. This criterion was not applied to the measurements performed in section~\ref{sec:bias}, as the number of bad pixels itself was evaluated.
	
	\section{Noise and bias optimisation}
	\label{sec:opti}
	Operation of the DePFET requires a set of static (Source, Drain, Back) and switched (Gate, Clear, Clear-Gate) voltages, designated as $V_\mathrm{DS}$, $V_\mathrm{BS}$, $V_\mathrm{GS,on/off}$, $V_\mathrm{CLS,on/off}$, and $V_\mathrm{CGS,on/off}$ (all referenced to the Source). The goal of operational optimisation is a bias configuration that yields a minimal ENC and, secondarily, a minimal $\mathrm{FWHM_{all}}$. As a starting point, well-known legacy operating conditions were used and adapted in broad terms to provide a sufficiently reliable basis for subsequent optimisation~\cite{sommerCharacterisationNewAthenaWFIs2026b}. Measurements were performed at a sensor temperature of approximately $220\,\mathrm{K}$ with a Clear pulse duration of $1.1\,\upmu\mathrm{s}$, a settling time of $2\,\upmu\mathrm{s}$ and an integration time of $4\,\upmu\mathrm{s}$. The sweep limits were constrained by the permissible voltage and current ratings of the Switcher and the Veritas. The former imposes $\lvert V_\mathrm{on} - V_\mathrm{off}\rvert \geq 7\,\mathrm{V}$ on all switched voltages, whereas the latter restricts $I_\mathrm{DS}$ to a maximum of $285\,\upmu\mathrm{A}$ due to the subtraction current limits. The measurements described below and the results obtained are summarised in table~\ref{tab:bias} and table~\ref{tab:optima}.
	
	\subsection{Peripheral bias and clear-process optimisation}
	\label{sec:bias}
	To rule out any considerable contribution to the overall noise and ensure robust operation, the backside voltage and clear-process relevant voltages were optimised prior to the $V_\mathrm{DS}$ and $V_\mathrm{GS,on}$ voltages, whose settings influence the intrinsic DePFET noise.
	
\begin{figure}[htbp]
		\centering 
		\includegraphics[width=0.55\textwidth,clip]{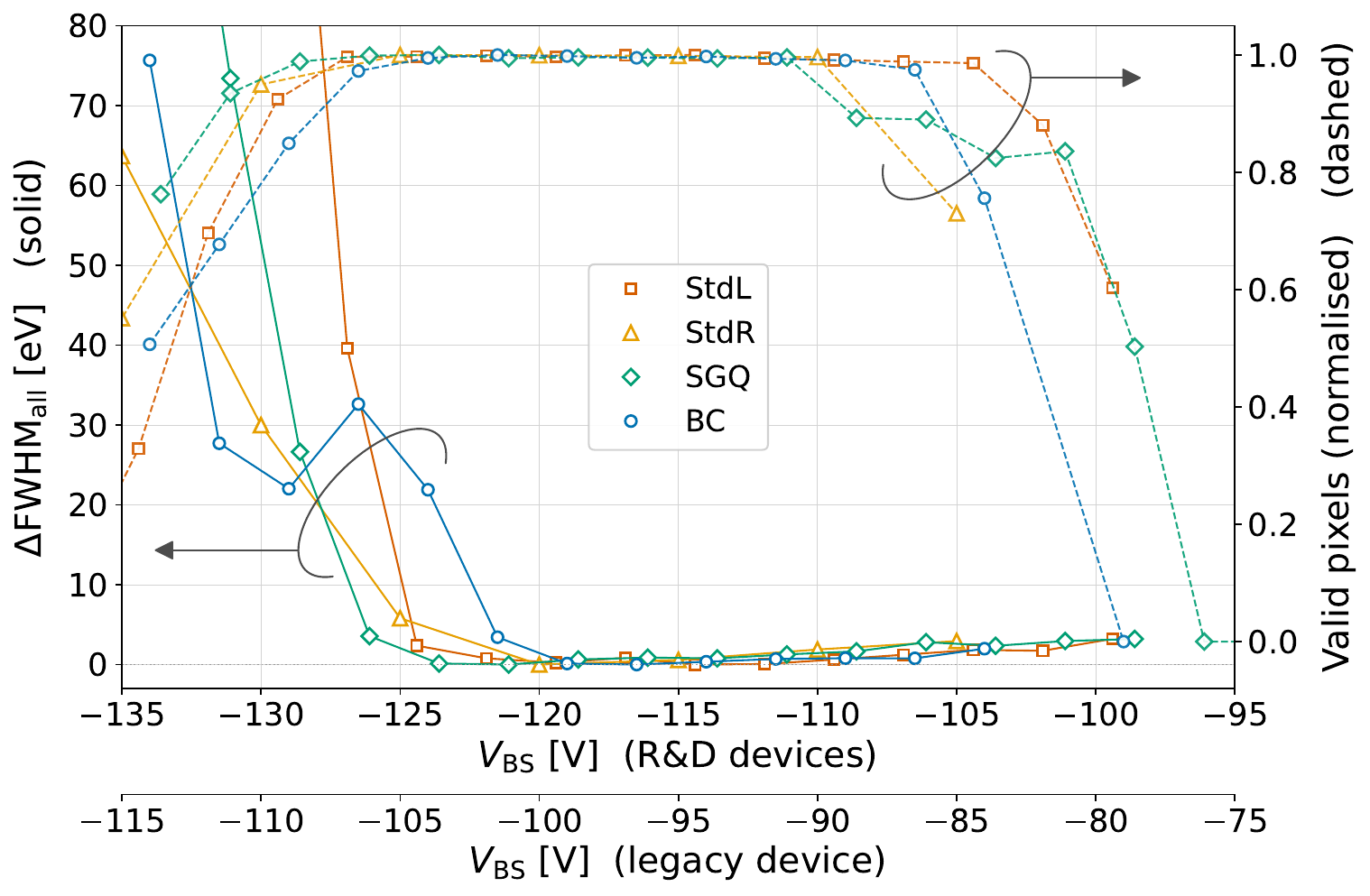}
		\caption{\label{fig:BacksideSweep} Results of the backside voltage ($V_\mathrm{BS}$) measurement for all four devices, showing the variation of $\mathrm{FWHM_{all}}$ and the normalised fraction of valid pixels, used as a proxy metric for the number of active pixels. The $V_\mathrm{BS}$ axis is plotted separately for the legacy device and the R\&D devices.}
	\end{figure}

	The backside voltage $V_\mathrm{BS}$ was made progressively more negative until the maximum number of pixels were sensitive and the lowest $\mathrm{FWHM_{all}}$ was obtained (figure~\ref{fig:BacksideSweep}). The latter represents the charge collection efficiency of the active pixels, and the sudden increase at low $V_\mathrm{BS}$ indicates charge losses to the clear. An additional headroom of $-5\,\mathrm{V}$ was reserved to accommodate subsequent sweeps that may shift the frontside potential. The legacy device reaches depletion $20\,\mathrm{V}$ earlier than the R\&D devices due to the different resistivity of the respective wafer lots. 
	
\begin{figure}[htbp]
		\centering 
		\includegraphics[width=0.49\textwidth,clip]{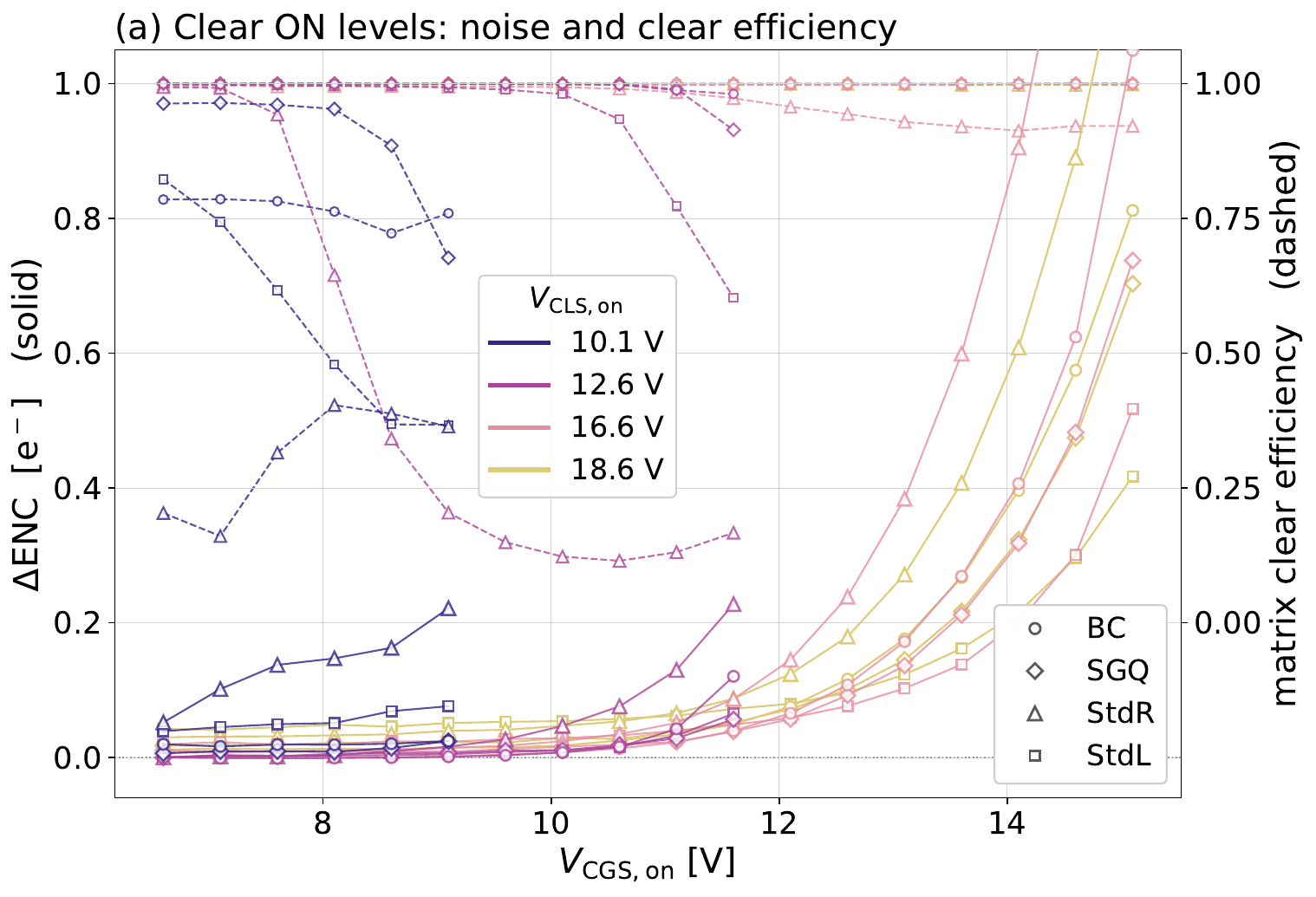}
		\includegraphics[width=0.49\textwidth,clip]{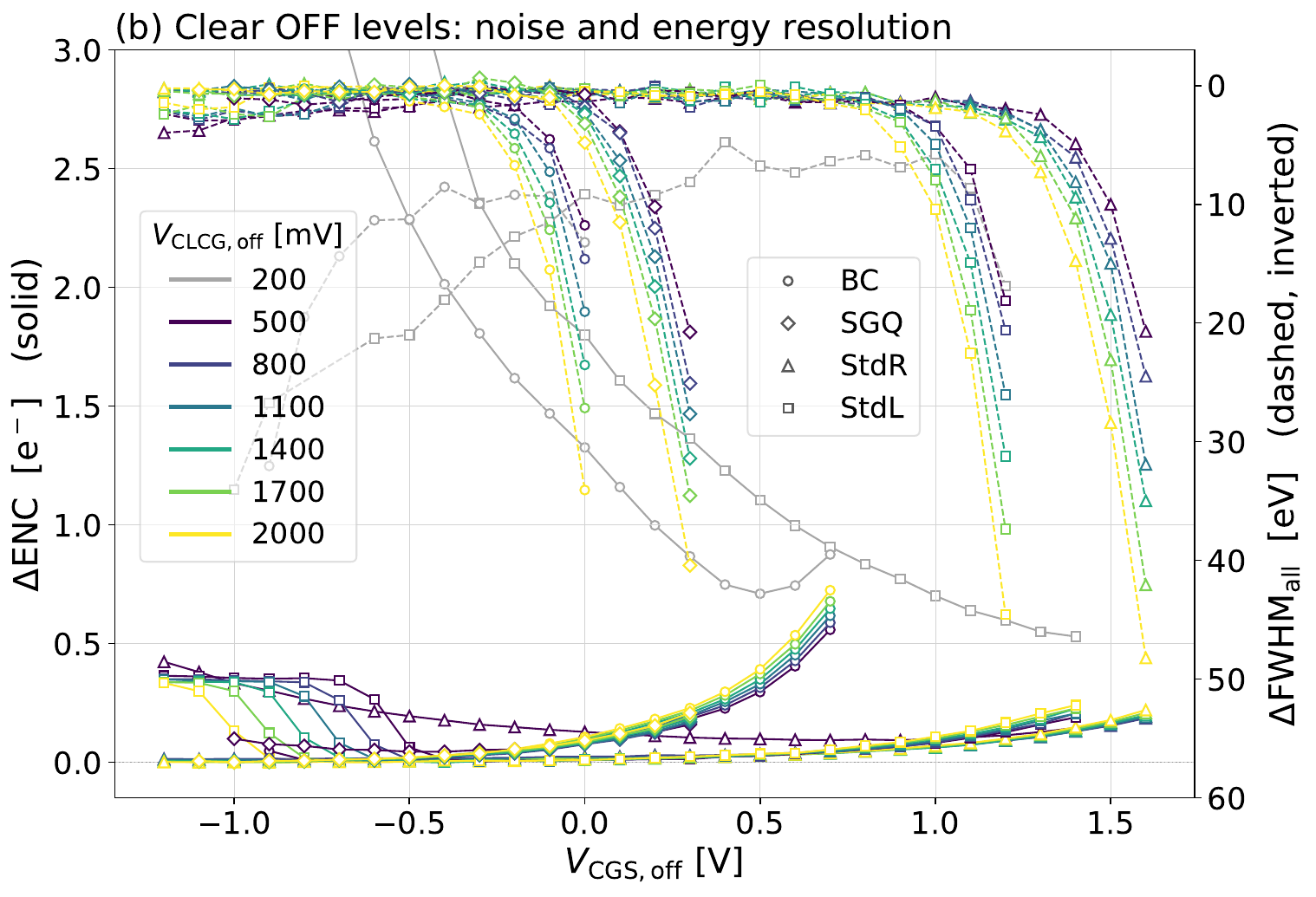}
		\caption{\label{fig:CLCGSsweeps} (a) Results of the Clear-Gate on-level sweep for four Clear settings, showing the variation of ENC and the matrix clear efficiency, which was derived from the bad pixels due to an insufficient clear and normalised to unity. (b) Results of the nested Clear/Clear-Gate off-level sweep. Variations in ENC and $\mathrm{FWHM_{all}}$ are shown relative to their respective optimum values. $V_\mathrm{CLS,off}$ was varied relative to $V_\mathrm{CGS,off}$, from $+2.0\,\mathrm{V}$ down to $+0.2\,\mathrm{V}$.}
	\end{figure}

	The on-levels of Clear and Clear-Gate were chosen sufficiently high to establish a potential gradient from the IG across the Clear-Gate to the Clear contact. For $V_\mathrm{CLS,on} \gtrsim 16\,\mathrm{V}$, both ENC and signatures of incomplete clear reach a combined minimum that is independent of $V_\mathrm{CGS,on}$ up to about $10\,\mathrm{V}$ (figure~\ref{fig:CLCGSsweeps} (a)). For higher $V_\mathrm{CGS,on}$, the clearing remains essentially complete, but the ENC rises. When $V_\mathrm{CGS,on}$ approaches $V_\mathrm{CLS,on}$, the clearing efficiency degrades, and with it the ENC. Therefore, a $V_\mathrm{CLS,on} = 17\,\mathrm{V}$ was selected, which does not compromise performance and allows complete clearing within shorter clear windows (section~\ref{sec:optimal}). The Clear-Gate on-level was set to $8\,\mathrm{V}$, within the flat region.
	
	The off-levels were optimised in a nested sweep on $\mathrm{FWHM_{all}}$ and ENC (figure~\ref{fig:CLCGSsweeps} (b)). Their joint operating window is bounded by three failure modes:
	\begin{enumerate}[label=(\arabic*), noitemsep, topsep=0pt, partopsep=0pt]
		\item If $V_\mathrm{CGS,off}$ is too negative (below approximately $-1\,\mathrm{V}$), inversion occurs beneath the Clear-Gate and, since this region encloses Source and Drain (figure~\ref{fig:depfetCut}, yellow structure), a parasitic channel is formed. 
		\item If $V_\mathrm{CLS,off}$ is too negative in combination with an insufficiently negative $V_\mathrm{CGS,off}$, the Clear-Gate barrier is weakened and allows back-injection of electrons from the Clear region into the IG. 
		\item If $V_\mathrm{CLS,off}$ is too positive, the Clear constitutes a competing sink to the IG, which increases the probability of signal-charge losses. In combination with a positive $V_\mathrm{CGS,off}$, no sufficient blocking off-state is established. 
	\end{enumerate}
	Based on these constraints, $V_\mathrm{CLS,off}$ was set to a slightly positive value of $+0.4\,\mathrm{V}$, such that the p-well barrier to the bulk is maintained (figure~\ref{fig:depfetCut}, right). The Clear-Gate off-level of $-0.4\,\mathrm{V}$ was found to be sufficient to provide a barrier without triggering failure mode (1). Apart from the backside voltage, the discussed bias parameters present a suitable compromise and were applied to all devices for simplicity.
	
	\begin{table}[htbp]
		\centering
		\caption{\label{tab:bias} Summary of peripheral bias configuration, common among the devices.}
		\smallskip
		\sbox{\biasTableBox}{
			\begin{tabular}{|ll|c|c|}
				\hline
				\multicolumn{2}{|c|}{} & \multicolumn{2}{c|}{Value [V]} \\
				\multicolumn{2}{|c|}{Voltage} & on & off \\
				\hline
				$V_\mathrm{CLS}$ & Clear & $+17$ & $+0.4$ \\
				$V_\mathrm{CGS}$ & Clear-Gate & $+8$ & $-0.4$ \\
				$V_\mathrm{BS}$ & Backside & \multicolumn{2}{c|}{$-95$\textsuperscript{(a)}, $-115$\textsuperscript{(b)}} \\
				\hline
			\end{tabular}
		} 
		\usebox{\biasTableBox}\par 
		\smallskip
		\begin{minipage}{\wd\biasTableBox}
			\footnotesize
			\textsuperscript{(a)} Legacy production, \quad \textsuperscript{(b)} R\&D production
		\end{minipage}
	\end{table}
	
	\subsection{Intrinsic bias-dependent noise and gain}
	\label{sec:bias2}
	The DePFET constitutes the first amplification stage in the readout chain and therefore represents the primary noise contribution during readout. Its bias-dependent noise behaviour was investigated by varying $V_\mathrm{GS,on}$ at several drain current levels, while $V_\mathrm{DS}$ follows from the bias conditions. This approach probes the accessible bias space both along and across lines of constant $I_\mathrm{DS}$, and since $g_q$ scales with $\sqrt{I_\mathrm{DS}}$ to first order (eq.~\eqref{eq:gq_eq2}), fixing $I_\mathrm{DS}$ separates the noise contributions that depend on the drain current from the remaining field-dependent effects. Moreover, it allows a comparison of the DUTs at equal $I_\mathrm{DS}$. The results for each device are shown in figures~\ref{fig:overviewGAIds} and~\ref{fig:detailsGAIds}, while the matrix scatter plots in figures~\ref{fig:matrixAthena}--\ref{fig:matrixSuper} (appendix~\ref{app:CCmatrix}) provide a more comprehensive overview, including $\mathrm{Noise_{out}}$, $\mathrm{FWHM_{all}}$ and the median $G_\mathrm{E}$. Relative changes in $G_\mathrm{E}$ can be attributed to the DePFET's internal amplification, since the gain of the downstream electronics remained unchanged, keeping the current-to-ADU conversion constant at $A = 0.0135(5)\,\mathrm{ADU/pA}$.
	
	\subsubsection{Noise and operating point optimisation}
	\label{sec:noise_opt}
\begin{figure}[htbp]
		\centering 
		\includegraphics[width=1\textwidth,clip]{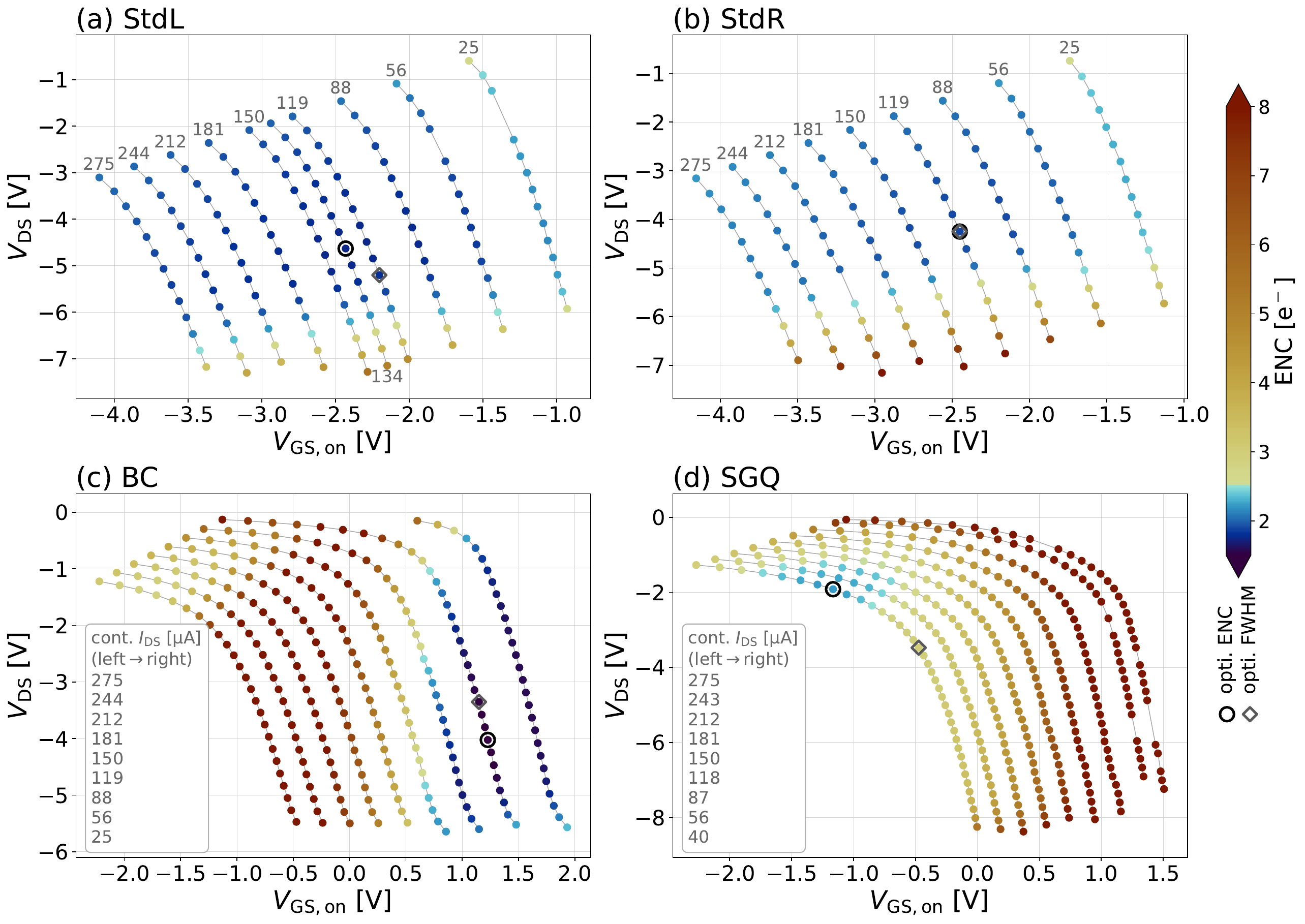}
		\caption{\label{fig:overviewGAIds} Overview of the DePFET transistor bias sweep results for the four DUTs in the $V_\mathrm{GS,on}$--$V_\mathrm{DS}$ plane. Each point represents one measured bias setting, coloured by its ENC on a diverging scale centred at $2.5\,\mathrm{e}^-$, separating the operating conditions of high noise (red) from low noise (blue). Lines connect settings of constant $I_\mathrm{DS}$, labelled in $\upmu\mathrm{A}$ at their ends or in the legends. Measurements were performed at a sensor temperature of $\approx 220\,\mathrm{K}$ and using an integration time of $4\,\upmu\mathrm{s}$. For clarity, only a subset of the measured data is presented. Complete sweep results are provided in appendix~\ref{app:CCmatrix}.}
	\end{figure}

	All devices exhibit a low-noise region, but its extent differs considerably (figure~\ref{fig:overviewGAIds} (a) to (d)). In the standard devices, the low-noise region extends broadly in both $V_\mathrm{DS}$ and $V_\mathrm{GS,on}$, whereas in \BC{} it is narrower and confined to positive $V_\mathrm{GS,on}$, with a second, albeit noisier, region at more negative $V_\mathrm{GS,on}$. The \SGQ{} device develops a low-noise region only towards the highest currents and small $\lvert V_\mathrm{DS}\rvert$.
	
	The optimal bias settings for each device in terms of minimal ENC are found within the low-noise region (figure~\ref{fig:overviewGAIds} (a) to (d), circle markers). For \SGQ{}, this optimum consequently lies at the boundary of the measured bias space. For the standard devices and \BC{}, the optima lie within the measurement limits.
	
\begin{figure}[htbp]
		\centering 
		\includegraphics[width=1\textwidth,clip]{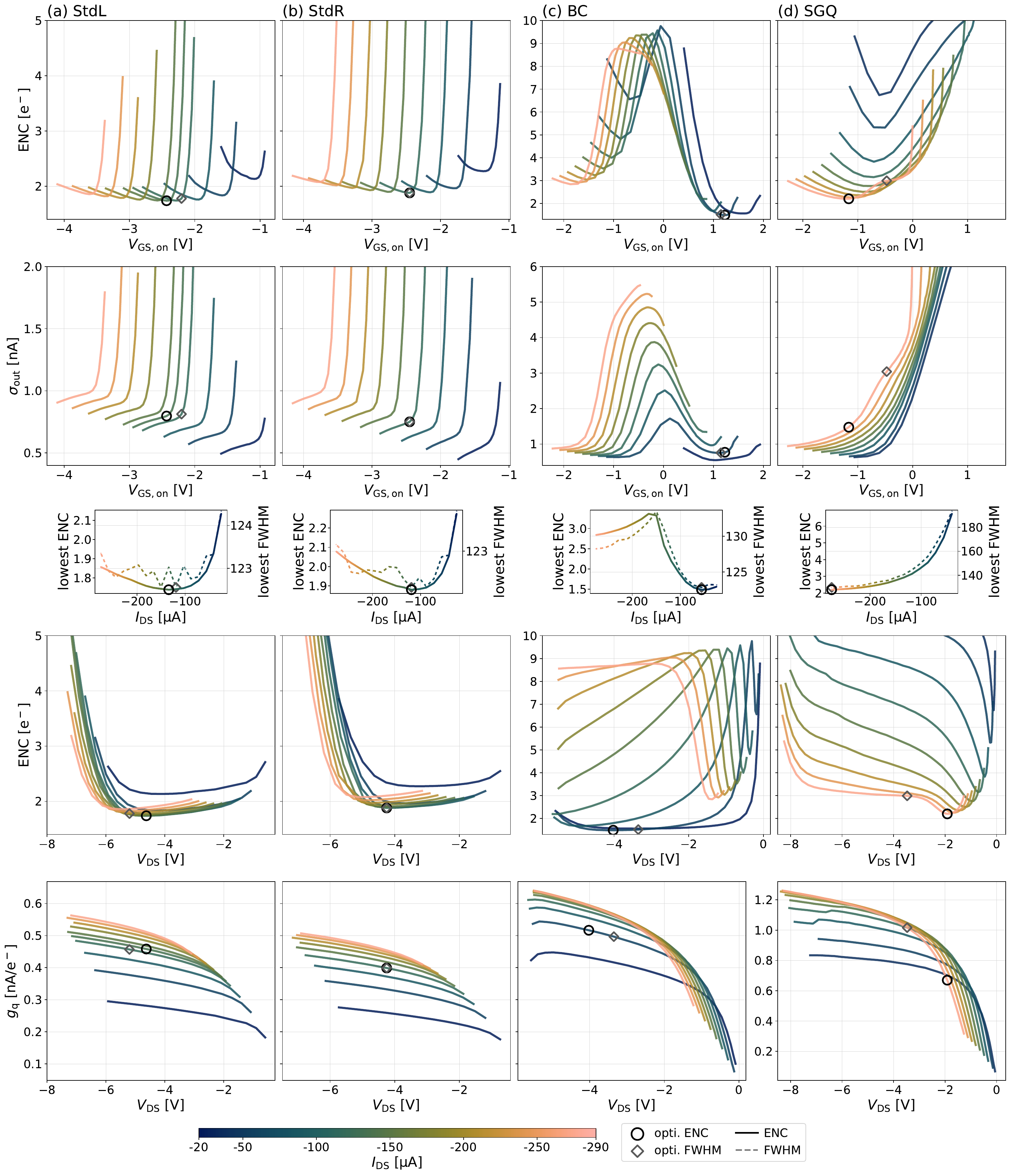}
		\caption{\label{fig:detailsGAIds} Results of the DePFET transistor bias sweep of figure~\ref{fig:overviewGAIds} replotted in detail. Columns are the DePFET devices (a) \StdL{}, (b) \StdR{}, (c) \BC{} and (d) \SGQ{}. ENC and $\sigma_\mathrm{out}$ as functions of $V_\mathrm{GS,on}$ (top two panel rows) and ENC as a function of $V_\mathrm{DS}$ (fourth panel row), at several constant drain currents (colour). The third panels show, for each $I_\mathrm{DS}$, the lowest achieved ENC and $\mathrm{FWHM_{all}}$. Bottom panel row depicts $g_q$ as a function of $V_\mathrm{DS}$, calculated from $G_\mathrm{E}$ using eq.~\eqref{eq:GE_to_gq}. The circle marks the ENC optimum across the measured bias space, and the diamond marks the $\mathrm{FWHM_{all}}$ optimum. For clarity, only a subset of the measured data is presented. Complete sweep results are provided in appendix~\ref{app:CCmatrix}.}
	\end{figure}

	Common to all devices is the behaviour of the charge gain (figure~\ref{fig:detailsGAIds}, $g_q(V_\mathrm{DS})$). At small $\lvert V_\mathrm{DS}\rvert$, where the DePFET transistor operates in the linear regime, $g_q$ rises steeply and continues to increase in the transition to saturation, but at a decreasing rate. Once saturation is reached at more negative $V_\mathrm{DS}$, $g_q$ increases steadily, but more gradually. This increase is consistent with channel length modulation (CLM)~\cite{enzChargeBasedMOSTransistor2006b}, to which a reduced effective IG capacitance at higher lateral field may also contribute in the standard devices and \BC{}.\footnote{These contributions cannot be conclusively separated with the available measurement data.} In \SGQ{}, the IG is already confined and, given its two-section channel nature, it is unclear whether CLM is the primary second-order effect.
	
	According to eq.~\eqref{eq:enc_sigma_gq} the ENC scales inversely with the charge gain ($\mathrm{ENC} \propto \sigma_\mathrm{out}/g_q$). In the linear regime, where $g_q$ is small, the ENC is correspondingly high, which limits the low-noise region of every device towards smaller $\lvert V_\mathrm{DS}\rvert$. In saturation, the ENC benefits from the gradual increase in $g_q$. In deep saturation, the ENC increases sharply (figure~\ref{fig:detailsGAIds}, $\mathrm{ENC}(V_\mathrm{DS})$), which is attributed to impact ionisation at the drain end of the channel.\footnote{For some current lines of \BC{} and \SGQ{}, the increase in ENC is not shown, since the number of bad pixels exceeded the threshold defined in section~\ref{subsec:settings_analysis}.} Further evidence is given in appendix~\ref{sec:impact_ionisation}. The onset of impact ionisation ($V_\mathrm{DS,II}$) becomes more negative with increasing current, since a higher current is associated with a larger saturation voltage $\lvert V_\mathrm{DS,sat}\rvert$ and thus a larger $\lvert V_\mathrm{DS}\rvert$ to reach the same peak lateral field. \SGQ{} stands out with $V_\mathrm{DS,II}$ about $2\text{--}3\,\mathrm{V}$ more negative than in the other devices (table~\ref{tab:optima}), which indirectly confirms the reduced field at the drain end predicted by simulations~\cite{bahrAdvancedDePFETConcepts2020}.
	For the standard devices and \BC{}, impact ionisation imposes another limit on the low-noise region, whereas the low-noise region of \SGQ{} is confined to the transition regime well below $\lvert V_\mathrm{DS,II} \rvert$.
	
	In the following, the noise behaviour per device is explored in greater detail.
	\emph{The standard devices} reach their lowest ENC at each constant current line close to the onset of impact ionisation due to the gradual increase in $g_q$. With increasing $\lvert I_\mathrm{DS}\rvert$, this lowest ENC initially improves, reaching the optimum before degrading again (figure~\ref{fig:detailsGAIds} (a, b), lowest ENC($I_\mathrm{DS}$)). This is because $g_q$ increases at a decreasing rate with $\lvert I_\mathrm{DS}\rvert$ (figure~\ref{fig:detailsGAIds} (a, b), $g_q$($V_\mathrm{DS}$), e.g. at $V_\mathrm{DS,II}$), whereas $\sigma_\mathrm{out}$ increases approximately linearly (figure~\ref{fig:detailsGAIds} (a, b), $\sigma_\mathrm{out}$($V_\mathrm{GS,on}$), points per current line before the upward slope). The optimum thus marks the balance point between the two, according to eq.~\eqref{eq:enc_sigma_gq}. Together with the limits set by the linear regime and impact ionisation, these constraints bound the low-noise region on all sides. The \StdR{} device exhibits an overall higher noise level than the \StdL{} device, which is attributed to inter-production variations.
	
	\emph{For the \BC{} device}, the low-noise region is additionally confined by $V_\mathrm{GS,on} \gtrsim 1\,\mathrm{V}$, where the device operates in buried-channel mode. This also restricts the low-noise region to currents below about $85\,\upmu\mathrm{A}$, since at higher currents the bias points at which the channel remains buried lie within the impact-ionisation regime (figure~\ref{fig:detailsGAIds} (c), $\mathrm{ENC}(V_\mathrm{DS})$ and $\mathrm{ENC}(V_\mathrm{GS,on})$). With decreasing $V_\mathrm{GS,on}$, a surface inversion layer forms, providing an additional current path for holes at the interface. Consequently, $\sigma_\mathrm{out}$, and with it the ENC, increases rapidly and reaches a maximum at $V_\mathrm{GS,on}$ between $-0.4\,\mathrm{V}$ and $0.2\,\mathrm{V}$ (figure~\ref{fig:detailsGAIds} (c), $\sigma_\mathrm{out}(V_\mathrm{GS,on})$). Its magnitude depends on the energy distribution of interface traps, their $V_\mathrm{GS,on}$ dependent occupancy, their capture/emission kinetics, which govern their contribution within the measurement bandwidth, and their electrostatic coupling to the channel~\cite{wilcoxLowtemperatureCharacterizationBuriedchannel1989, christenssonLowFrequencyNoise1968}. Beyond the maximum, the decrease in $\sigma_\mathrm{out}$ is consistent with interface trap states reaching saturated occupancy. The resulting local ENC minimum constitutes the second, noisier region (figure~\ref{fig:detailsGAIds} (c), $\mathrm{ENC}(V_\mathrm{GS,on})$). There, the ENC depends strongly on $I_\mathrm{DS}$, since decreasing $\lvert I_\mathrm{DS}\rvert$ lowers $\lvert V_\mathrm{DS}\rvert$ and moves the operating point from the transition into the linear regime (figure~\ref{fig:detailsGAIds} (c), $\mathrm{ENC}(V_\mathrm{DS})$, ENC dip depth increases).
	
	\emph{For the \SGQ{} device}, the ENC does not improve as $V_\mathrm{DS}$ becomes more negative for any given $I_\mathrm{DS}$, unlike in the other devices, since the increase in $\sigma_\mathrm{out}$ outweighs the gradual increase in $g_q$ noted above (figure~\ref{fig:detailsGAIds} (d), $\sigma_\mathrm{out}(V_\mathrm{GS,on})$). Furthermore, at a constant $V_\mathrm{DS}$, the decrease in ENC with increasing $\lvert I_\mathrm{DS}\rvert$ exceeds the scaling expected for a thermal noise contribution (eq.~\eqref{eq:enc_thermal_scaling}; figure~\ref{fig:detailsGAIds} (d), ENC($V_\mathrm{DS}$)). Both observations point to an excess-noise contribution that increases towards more positive $V_\mathrm{GS,on}$ (figure~\ref{fig:detailsGAIds} (d), $\sigma_\mathrm{out}(V_\mathrm{GS,on})$), since both a more negative $V_\mathrm{DS}$ at constant current and a lower current at constant $V_\mathrm{DS}$ shift the operating point in this direction. Below approximately $150\,\upmu\mathrm{A}$, the increasing $g_q$ additionally lowers the thermal contribution, whereas above this current, $g_q$ becomes less dependent on $\lvert I_\mathrm{DS} \rvert$, and any further improvement in ENC is largely attributable to the reduction of this excess noise. This improvement, however, diminishes with increasing current (figure~\ref{fig:detailsGAIds} (d), lowest ENC($I_\mathrm{DS}$)), and currents beyond the measurement limit are therefore not expected to reduce the ENC considerably. At small $\lvert V_\mathrm{DS}\rvert$, within the transition regime, a noise valley appears which exhibits the lowest ENC for each imposed $I_\mathrm{DS}$ and, towards the highest currents, forms the low-noise region (figure~\ref{fig:detailsGAIds} (d), ENC($V_\mathrm{DS}$) and lowest ENC($I_\mathrm{DS}$)). 
	
	\begin{table}[htbp]
		\centering
		\caption{\label{tab:optima} Optimal operating points of the four DePFET matrices, selected for minimal ENC across the transistor bias space. $V_\mathrm{DS}$ follows from the $V_\mathrm{GS,on}$ at the given $I_\mathrm{DS}$. $V_\mathrm{DS,II}$ marks the onset of impact ionisation at the stated bias $I_\mathrm{DS}$. $\mathrm{FWHM_{all}}$ and $\mathrm{FWHM_{sng}}$ are reported at the ENC optimum. Their standard deviation ($\hat\sigma_\mathrm{SD}$) was estimated from the second differences of consecutive measurements of constant $I_\mathrm{DS}$ sweeps. Applying the same estimation method to ENC, but with the median absolute deviation, yields a $\hat\sigma_\mathrm{MAD}$ below $0.01\,\mathrm{e}^-$ over all devices.}
		\smallskip
		\begin{tabular}{|l|cccc|ccc|c|}
			\hline
			& $\lvert I_\mathrm{DS} \rvert$ & $V_\mathrm{DS}$ & $V_\mathrm{GS,on}$ & $V_\mathrm{GS,off}$ & ENC
			& $\mathrm{FWHM_{all}}$ & $\mathrm{FWHM_{sng}}$
			& $V_\mathrm{DS,II}$ \\
			Device & [$\upmu\mathrm{A}$] & [V] & [V] & [V] & [$\mathrm{e}^-$] & [eV] & [eV] & [V] \\
			\hline
			\StdL{} & $134$ & $-4.63$ & $-2.43$ & $V_\mathrm{GS,on} + 7$ & $1.74$ & $123.0(3)$ & $120.7(5)$ & $-5.2$ \\ 
			\StdR{} & $119$ & $-4.25$ & $-2.45$ & $V_\mathrm{GS,on} + 7$ & $1.88$ & $123.7(4)$ & $121.7(5)$ & $-5.2$ \\ 
			\BC{} & $56$ & $-4.02$ & $1.23$ & $V_\mathrm{GS,on} + 7$ & $1.49$ & $123.2(3)$ & $121.2(5)$ & $-4.5$ \\ 
			\SGQ{}\textsuperscript{(a)} & $275$ & $-1.92$ & $-1.17$ & $V_\mathrm{GS,on} + 7$ & $2.20$ & $133.2(7)$ & $123.2(10)$ & $-7.3$ \\ 
			\hline
		\end{tabular}
		\smallskip
		\begin{minipage}{0.9\textwidth}
			\footnotesize
			\textsuperscript{(a)} The ENC optimum occurs within the transition to the linear regime, whereas the $\mathrm{FWHM_{all}}$ optimum of $131\,\mathrm{eV}$ is at the onset.
		\end{minipage}
	\end{table}

	The energy resolution largely follows the ENC for the standard devices and \BC{}. For \StdR{}, the $\mathrm{FWHM_{all}}$ optimum coincides with the ENC optimum, whereas the slightly shifted $\mathrm{FWHM_{all}}$ optima of \StdL{} and \BC{} are only apparent, given the scatter in $\mathrm{FWHM_{all}}$ (table~\ref{tab:optima}). For \SGQ{}, however, $\mathrm{FWHM_{all}}$ increases within the noise valley. It reaches its optimum of $131\,\mathrm{eV}$ at the onset of saturation. This increase in $\mathrm{FWHM_{all}}$ is not reflected in $\mathrm{FWHM_{sng}}$, which follows the noise and remains comparatively small (table~\ref{tab:optima}), nor can the combined noise of multi-pixel events account for it. Its origin remains open.
	
	\begin{figure}[htbp]
		\centering 
		\includegraphics[width=1\textwidth,clip]{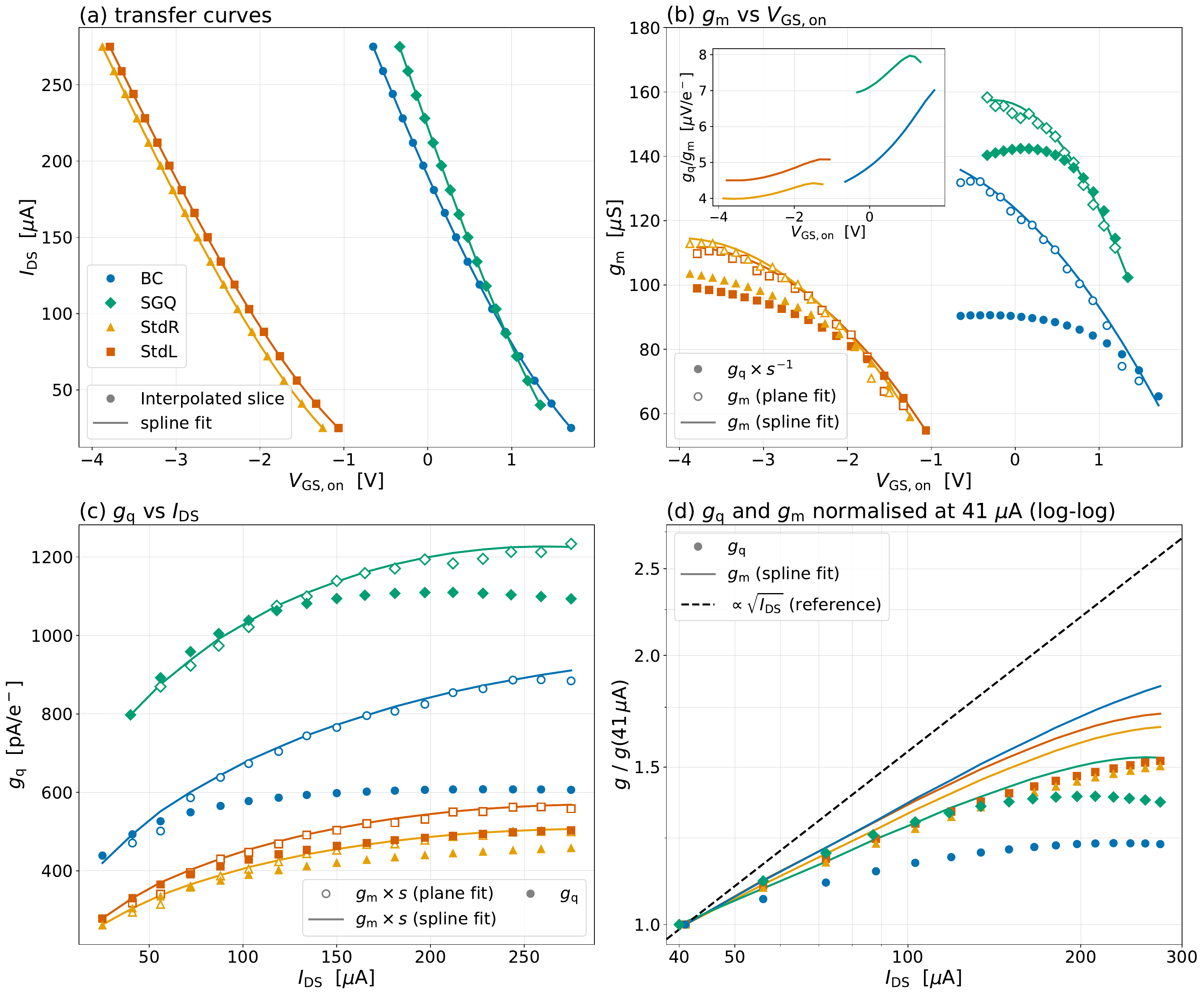}
		\caption{\label{fig:gqgm} Charge gain and transconductance of the four DUTs, derived from the bias sweep of figure~\ref{fig:overviewGAIds} at a common $V_\mathrm{DS} = -4.4\,\mathrm{V}$ locus. (a) Transfer curves, with the spline fitted to the interpolated points. (b) Transconductance as a function of $V_\mathrm{GS,on}$ and (c) charge gain as a function of $I_\mathrm{DS}$. Where $g_q$ and $g_\mathrm{m}$ diverge, the internal-gate coupling changes. In both plots (b, c), the respective other quantity is scaled by the factor $s = g_q/g_\mathrm{m}$, taken at $41\,\upmu\mathrm{A}$. The inset in (b) shows $g_q/g_\mathrm{m}$ (i.e. $q\,f/C_\mathrm{IG}$), over the full bias range. (d) Both quantities normalised to their value at $41\,\upmu\mathrm{A}$, together with the ideal $\sqrt{I_\mathrm{DS}}$ scaling as a reference. Transconductance was derived by a small-signal plane fit directly to the swept parameter space (open markers) and, as a smooth overlay (lines), by differentiating the spline fitted to the transfer curves.}
	\end{figure}

	\subsubsection{Charge gain and transconductance}	
	\label{sec:gain}
	The transconductance describes the small-signal gain of the DePFET transistor, whereas the charge gain additionally accounts for the coupling between the internal gate and the channel (eq.~\eqref{eq:gq_gm}). Their joint analysis separates variations in transistor performance from variations of the internal-gate coupling. A reliable comparison among all DUTs requires a common $V_\mathrm{DS}$ at which the devices are operated in saturation and below their onset of impact ionisation (table~\ref{tab:optima}). Since $V_\mathrm{DS}$ follows from the bias conditions in the constant $I_\mathrm{DS}$ sweeps, the transfer curves and corresponding $g_q$ were obtained by linear interpolation to a locus at $V_\mathrm{DS} = -4.4\,\mathrm{V}$ (figure~\ref{fig:gqgm} (a)). The transconductance was derived from a local plane fit applied directly to the measurement points neighbouring the locus in the $V_\mathrm{GS,on}$--$V_\mathrm{DS}$ plane, whereas the smooth overlaid curve was obtained by differentiating the spline fitted to the transfer curves (figure~\ref{fig:gqgm} (b)).\footnote{Plane fit using $I_\mathrm{DS} \approx g_\mathrm{m} V_\mathrm{GS,on} + g_\mathrm{ds} V_\mathrm{DS} + c$ and with an evaluation window around each point ranging $\pm 35\,\upmu\mathrm{A}$ in $I_\mathrm{DS}$ and $\pm 1.2\,\mathrm{V}$ in $V_\mathrm{DS}$.}
	
	\BC{} and \SGQ{} require a considerably more positive $V_\mathrm{GS,on}$ than the standard devices to reach the same $I_\mathrm{DS}$, reflecting their shift in threshold voltage caused by the larger gate-channel separation and the higher shallow p-channel doping, respectively (figure~\ref{fig:gqgm} (a)).
	
	Among all four devices, the transconductance increases more gradually than the ideal $\sqrt{I_\mathrm{DS}}$ dependence. Over the measured current range, the extracted exponents for $g_\mathrm{m}$ range from $0.22$ to $0.32$ (figure~\ref{fig:gqgm} (d)). This deviation can be attributed to vertical field mobility degradation, source/drain series resistances as well as the onset of velocity saturation.
	
	Both novel devices attain a higher charge gain than the standard devices (figure~\ref{fig:gqgm} (c)). At $150\,\upmu\mathrm{A}$, \SGQ{} reaches $1094\,\mathrm{pA/e^-}$ and \BC{} $598\,\mathrm{pA/e^-}$, an improvement by factors of $2.6$ and $1.4$ over \StdR{}, respectively. 
	For \BC{}, this mainly reflects a higher internal-gate coupling (figure~\ref{fig:gqgm} (b), inset), since the buried channel lowers $C_\mathrm{IG}$ (eq.~\eqref{eq:CG}). The generally higher $g_\mathrm{m}$, despite the lower gate-channel capacitance, additionally raises $g_q$ and is consistent with a higher charge carrier mobility observed in buried channels~\cite{huDesignTradeoffsSurface1985b}. The improvement in \SGQ{} is close to the factor of three predicted by simulations~\cite{stadlerFirstInsightsSupergq2024b} and stems from both a higher internal-gate coupling and a considerably larger transconductance (figure~\ref{fig:gqgm} (b)). Above $150\,\upmu\mathrm{A}$, $g_q$ becomes increasingly less dependent on $\lvert I_\mathrm{DS} \rvert$ in all devices, whereas $g_\mathrm{m}$ continues to rise, indicating a loss in coupling efficiency, since $C_\mathrm{IG}$ is fixed to first order in the standard and super-$g_q$ devices. This divergence is more pronounced in the \BC{} device and occurs at lower $\lvert I_\mathrm{DS} \rvert$, concomitant with the increase in noise at the corresponding $V_\mathrm{GS,on}$ mentioned previously (section~\ref{sec:noise_opt}). Here, the additional hole current path at the interface displaces the charge centroid of the channel towards the external gate. In other words, $x_\mathrm{s}$ gets smaller and $C_\mathrm{IG}$ rises again, thereby further weakening the internal-gate coupling compared to the other variants~\cite{wilcoxLowtemperatureCharacterizationBuriedchannel1989, radekaSignalProcessingParticle2020}.

	\section{Noise and temperature dependence}
	\label{sec:temperature}
\begin{figure}[htbp]
		\centering
		\includegraphics[width=1\textwidth,clip]{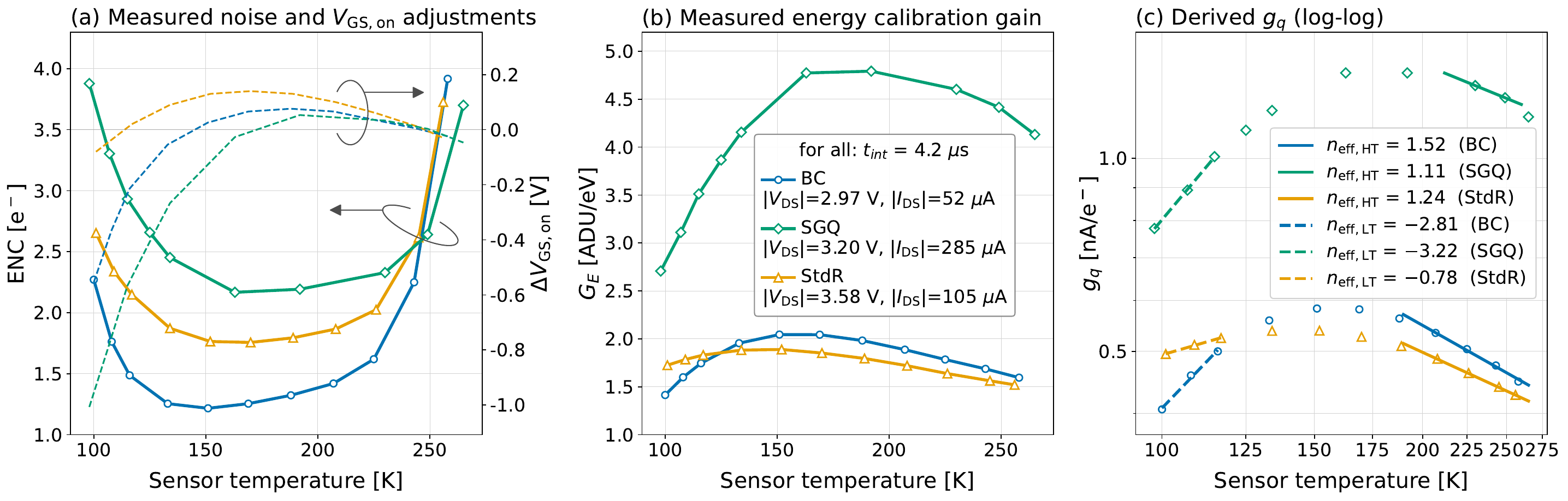}
		\caption{\label{fig:tempSweep} Sensor performance of \BC{}, \StdR{}, and \SGQ{} over a wide temperature range under the indicated bias conditions. Temperature dependence of $\mathrm{ENC}$ as well as the $V_\mathrm{GS,on}$ adjustments relative to $250\,\mathrm{K}$, required to maintain a constant current (a) and energy calibration gain $G_\mathrm{E}$ (b). Log-log scaled $g_q$ and extracted effective temperature exponents $n_\mathrm{eff,HT}$ and $n_\mathrm{eff,LT}$ (c), from linear fits over the high- and low-temperature ranges of $190\text{--}260\,\mathrm{K}$ and $100\text{--}120\,\mathrm{K}$, respectively (\SGQ{} indicative only).}
	\end{figure}

	The noise behaviour as a function of temperature was investigated for the \BC{}, \StdR{}, and \SGQ{} devices over a wide temperature range.\footnote{A detailed interpretation of the SGQ results requires a dedicated analysis of the two channel sections and the excess noise and is beyond the scope of this paper.} The bias settings were chosen according to tables~\ref{tab:bias} and~\ref{tab:optima}, with $V_\mathrm{DS}$ tuned to avoid impact ionisation and to balance $\mathrm{FWHM_{all}}$/ENC performance against sweep stability. $V_\mathrm{GS,on}$, on the other hand, was adjusted in order to maintain a constant $I_\mathrm{DS}$. The bias settings and measurement results are shown in figure~\ref{fig:tempSweep}. 
	
	The noise initially decreases as the temperature drops. Below $230\,\mathrm{K}$, the bulk leakage shot noise that prevailed up to this point contributes only marginally (figure~\ref{fig:tempSweep} (a)). With further cooling, the noise continues to improve more gradually as thermal noise and its weaker temperature dependence predominate, until a minimum is reached at around $155\,\mathrm{K}$ across all three devices, below which the noise increases again. There, increasingly larger $V_\mathrm{GS,on}$ adjustments were needed for all three devices (figure~\ref{fig:tempSweep} (a)), which suggests that incomplete ionisation (i.e.\ onset of carrier freeze-out) of the implanted p-channel dopants emerges and that the resulting mean carrier deficit must be compensated for~\cite{morinElectricalPropertiesSilicon1954}. 
	
	In the following, $g_q$ is examined to assess whether the noise increase below the minimum stems from the thermal noise, which scales with $\sqrt{T/g_q}$ (eq.~\eqref{eq:enc_thermal_scaling}). The measured $G_\mathrm{E}$ carries the temperature dependence of both $g_q$ and $w(T)$ (figure~\ref{fig:tempSweep} (b)). Since the readout settings remained unchanged and the electronics were held at a fixed temperature, $A$ is constant. Correcting $G_\mathrm{E}$ for $w(T)$ using eq.~\eqref{eq:GE_to_gq} therefore returns $g_q$. At a constant $I_\mathrm{DS}$, $g_q$ depends on temperature $T$ only through the effective mobility, according to eq.~\eqref{eq:gq_eq2} and the empirical power law such that 
	\begin{equation}
		\label{eq:mobility_gq}
		\mu(T) = \mu_0 \left( \frac{T}{T_0} \right)^{-n_\mathrm{eff}} \quad \Rightarrow \quad \mu \propto T^{-n_\mathrm{eff}} \quad \Rightarrow \quad g_q \propto T^{-n_\mathrm{eff}/2},
	\end{equation}
	with the reference temperature $T_0$, reference mobility $\mu_0$ at $T_0$, and the effective temperature exponent $n_\mathrm{eff}$~\cite{klaassenUnifiedMobilityModel1992}.\footnote{A temperature dependence of the internal-gate coupling is assumed to be negligible.}
	
	The charge gain rises with decreasing temperature down to $\approx 160\,\mathrm{K}$ and falls below it (figure~\ref{fig:tempSweep} (c)). Above the maximum the mobility is limited by phonon scattering, which decreases on cooling, and below it by ionised impurity scattering, which increases on cooling~\cite{klaassenUnifiedMobilityModel1992,kimataLowTemperatureCharacteristicsBuriedChannel1983}. Here, the effective mobility also includes the effect of the incomplete ionisation indicated by the $V_\mathrm{GS,on}$ adjustments, and separating both contributions requires further study to determine the density of mobile holes in the channel~\cite{wilcoxLowtemperatureCharacterizationBuriedchannel1989}. The $g_q$ maximum appears for the \SGQ{} device about $15\,\mathrm{K}$ higher, in line with its increased shallow p-channel doping dose~\cite{kimataLowTemperatureCharacteristicsBuriedChannel1983}. Rearranging eq.~\eqref{eq:mobility_gq} gives
	\begin{equation}
		\label{eq:mobility_exponent_meas}
		n_\mathrm{eff} = -2\,\frac{\mathrm{d}\log g_q}{\mathrm{d}\log T}.
	\end{equation}
	Thus, the effective temperature exponents $n_\mathrm{eff,HT}$ and $n_\mathrm{eff,LT}$ can be obtained from linear fits on a log-log scale over the high- and low-temperature ranges of $190\text{--}260\,\mathrm{K}$ and $100\text{--}120\,\mathrm{K}$, respectively (figure~\ref{fig:tempSweep} (c)). The larger $n_\mathrm{eff,HT}$ and $\lvert n_\mathrm{eff,LT} \rvert$ of \BC{} are consistent with a more bulk-like transport, unlike in \StdR{}, where the less temperature-dependent surface-roughness scattering and Coulomb scattering by oxide and interface charges play a more significant role~\cite{chengEffectCoulombScattering1974}.
	
	With this temperature dependence of $g_q$, the changes in thermal noise ($S_{v,\mathrm{th}}$) below its maximum are minor, since temperature enters the noise directly and through the declining $g_q$ in opposite directions~\cite{lutzSemiconductorRadiationDetectors2008}. Using eq.~\eqref{eq:gq_gm}, $\mathrm{ENC_{th}}$ can be expressed as
	\begin{equation}
		\label{eq:enc_thermal_scaling}
		S_{v,\mathrm{th}} = \frac{4k_\mathrm{B}T\gamma}{g_\mathrm{m}} \quad \Rightarrow \quad \mathrm{ENC_{th}} \propto \sqrt{T/g_q}\,.
	\end{equation}
	The expected $\mathrm{ENC_{th}}$ continues to fall by factors of $0.95$ (\BC{}) and of $0.82$ (\StdR{}) from $160\,\mathrm{K}$ down to $100\,\mathrm{K}$. However, the measured $\mathrm{ENC}$ rises by factors of $1.8$ and $1.6$, respectively, over the same temperature range. This indicates an additional noise source, presumably generation-recombination noise originating from the partially ionised channel dopants~\cite{arnaboldiTemperatureCharacterizationDeep2004,wilcoxLowtemperatureCharacterizationBuriedchannel1989,kandiahLimitsResolutionCharge1993a}.
	
	\section{Noise and integration time}
	\label{sec:optimal}
	The noise performance of all four devices was investigated with respect to the integration time $t_\mathrm{int}$. According to the time-variant filter formalism of Gatti et al.~\cite{gattiSuboptimalFiltering11990}, increasing $t_\mathrm{int}$ progressively suppresses the series white noise (thermal) component and, assuming negligible parallel white (bulk-leakage) noise at a sensor temperature of $158\,\mathrm{K}$, the total noise approaches its theoretical minimum. In this case, $1/f^{\alpha}$ (flicker) noise emerges as the dominant term.
	
	Measurements were performed by varying $t_\mathrm{int}$ from $0.4\,\upmu\mathrm{s}$ to $9\,\upmu\mathrm{s}$, where the upper limit was given by the dynamic range of the Veritas integration stage. Owing to its higher internal amplification, the \SGQ{} variant reached this constraint at $t_\mathrm{int} = 5\,\upmu\mathrm{s}$. To ensure comparability across the integration time sweep, the frame cycle time was held constant. At integration times below $9\,\upmu\mathrm{s}$, the reduced process time per row was compensated by an additional dwell time at the beginning of each frame cycle. Consequently, for each frame the same average number of events was captured and the surface- and residual bulk-leakage noise contributions between consecutive readouts remained constant. Furthermore, the flat-top time was optimised to $1.2\,\upmu\mathrm{s}$, with a settling time of $0.5\,\upmu\mathrm{s}$ and a Clear pulse duration of $0.7\,\upmu\mathrm{s}$ without causing an incomplete clear. At $t_\mathrm{int} = 9\,\upmu\mathrm{s}$, this results in a readout time of $19.2\,\upmu\mathrm{s}$ per row (i.e. per pixel).
	
	\begin{figure}[htbp]
		\centering 
		\includegraphics[width=1\textwidth,clip]{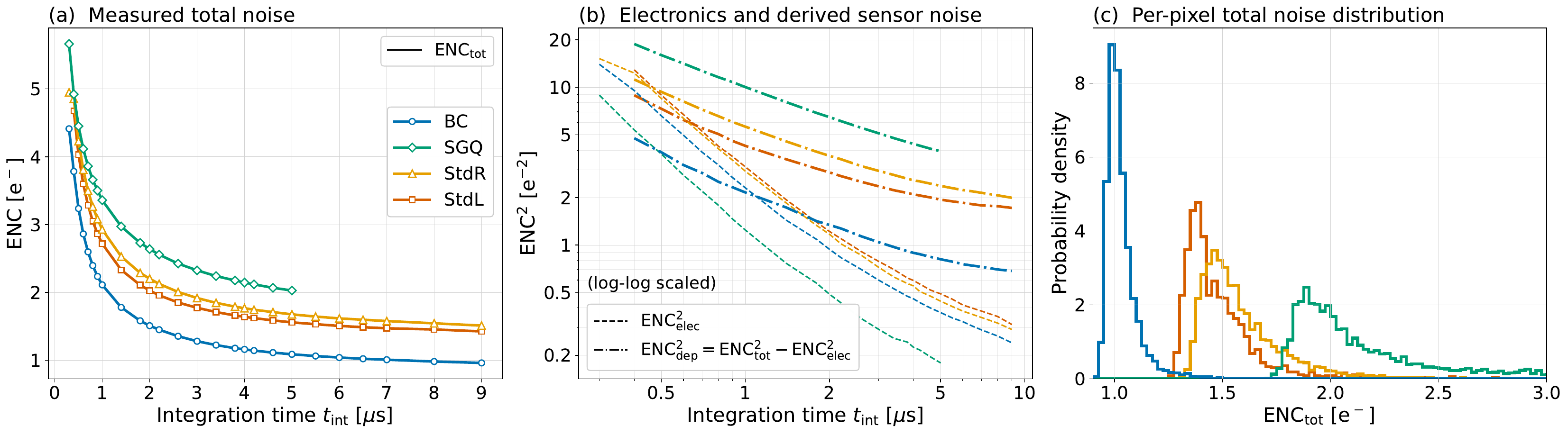}
		\caption{\label{fig:IntTimeSweep} Integration-time dependence of the noise at a sensor temperature of $158\,\mathrm{K}$ and the bias settings of table~\ref{tab:bias_enc_summary}. (a) Total noise. The upper end of the range is set by the dynamic range of the Veritas integration stage, which \SGQ{} reaches at $5\,\upmu\mathrm{s}$. (b) $\mathrm{ENC_{elec}}$, measured at $I_\mathrm{SSS} = 25\,\upmu\mathrm{A}$ per channel with the sensor turned off. The resulting $\mathrm{ENC_{dep}}$ is an upper estimate. (c) Pixel noise distributions, normalised as probability densities, at the respective longest measured integration time per device.}
	\end{figure}
	
	Moreover, the electronics noise of the readout chain from the Veritas to the ADC was examined. Frames were acquired in which the sensor matrix was turned off by setting $V_\mathrm{DS} = 0\,\mathrm{V}$ and $V_\mathrm{GS,on}$ high. The subtraction current $I_\mathrm{SSS}$ of the Veritas was set to $25\,\upmu\mathrm{A}$ per channel to avoid clipping of the I2V stage. Using the electronics $\mathrm{Noise_{out}}$, the measured $G_\mathrm{E}$ from the illuminated frames, and a pair generation energy of~$w(158\,\mathrm{K}) = 3.73\,\mathrm{eV}$, the electronics noise in ADU was converted into the input-referred noise ($\mathrm{ENC_{elec}}$). The $\mathrm{ENC_{elec}}$ results differ between the detector assemblies only through their corresponding charge gain (figure~\ref{fig:IntTimeSweep} (b)). Re-referred to the readout input using the respective $g_q$, the four assemblies agree in noise at about $375\,\mathrm{pA}$ ($5\,\upmu\mathrm{s}$) and $305\,\mathrm{pA}$ ($9\,\upmu\mathrm{s}$).
	Measurements at low $I_\mathrm{SSS}$, as well as dedicated measurements at higher $I_\mathrm{SSS}$ using a detector assembly without a sensor, which enabled the use of the Veritas Test-IN, revealed that the electronics noise increases monotonically with $I_\mathrm{SSS}$. However, $\mathrm{ENC_{elec}}$ was determined at an $I_\mathrm{SSS}$ not matching the respective device $I_\mathrm{DS}$ and therefore represents a lower estimate. Consequently, the upper estimate of the DePFET noise contribution $\mathrm{ENC_{dep}}$ to $\mathrm{ENC_{tot}}$ can be isolated by
	\begin{equation}
	\label{eq:enc_depfet}
	\mathrm{ENC_{dep}}
	=
	\sqrt{\mathrm{ENC_{tot}}^2
	- \mathrm{ENC_{elec}}^2}.
	\end{equation}
	
	The results obtained at the longest measured integration times of $5\,\upmu\mathrm{s}$ (\SGQ{}) and $9\,\upmu\mathrm{s}$ (other variants), as well as the operating settings, which were re-evaluated for minimal ENC and adjusted where necessary, are summarised in table~\ref{tab:bias_enc_summary}. 
	
	Among all variants, \BC{} has the lowest total noise, reaching $0.96\,\mathrm{e}^-$, a $36\%$ reduction over \StdR{}, to which both a lower $\sigma_\mathrm{out}$ and a higher $g_q$ contribute, by factors of $0.72$ and $1.14$ respectively. The improvement in $\sigma_\mathrm{out}$ persists even at the longest integration time, where the thermal contribution is increasingly filtered out, which points to a reduced $1/f$ contribution. \BC{} reaches this noise level at less than half the power consumption during readout compared to standard devices.\footnote{Estimated as $P = I_\mathrm{DS}\times V_\mathrm{DS}$ at the operating points in table~\ref{tab:bias_enc_summary}: $0.22\,\mathrm{mW}$ (\BC{}), $0.51\,\mathrm{mW}$ (\StdR{}) and $0.65\,\mathrm{mW}$ (\StdL{}).} The \SGQ{} device, on the other hand, yields an $\mathrm{ENC_{tot}}$ of $2.03\,\mathrm{e}^-$, which is about $21\%$ higher than that of \StdR{} at the same integration time. As its optimum lies within the transition regime (section~\ref{sec:noise_opt}), only part of its charge gain is realised. The remaining $g_q$ advantage over \StdR{}, a factor of $1.50$, is outweighed by a $1.82$ times higher $\sigma_\mathrm{out}$.
	
	The relative contribution of $\mathrm{ENC_{elec}}$ to $\mathrm{ENC_{tot}}$ for all detector assemblies decreases monotonically as $t_\mathrm{int}$ increases, and $\mathrm{ENC_{dep}}$ starts to dominate the noise budget (figure~\ref{fig:IntTimeSweep} (b)). $\mathrm{ENC_{dep}}$ does not allow a direct comparison of the sensors, given their different drain currents. However, even as an upper estimate, it shows that the electronics noise is not negligible. For example, in the case of \BC{}, it accounts for at least a quarter of the total noise variance at the longest integration time, and for about half at $t_\mathrm{int} = 1\,\upmu\mathrm{s}$ (figure~\ref{fig:IntTimeSweep} (b)). Higher Veritas gain settings can reduce the relative contribution of noise sources further down the readout chain, but only to some extent and at the cost of dynamic range (appendix~\ref{app:GainExplo})~\cite{treberspurgAchievableNoisePerformance2018a}.
		\begin{table}[tb]
		\centering
		\caption{Noise at integration times of $5\,\upmu\mathrm{s}$ and $9\,\upmu\mathrm{s}$, at the given bias settings and a sensor temperature of $158\,\mathrm{K}$. For \SGQ{}, the dynamic range of the Veritas limits the integration time to $5\,\upmu\mathrm{s}$.}
		\label{tab:bias_enc_summary}
		\smallskip
		\sbox{\biasenctable}{
			\resizebox{\columnwidth}{!}{
				\begin{tabular}{|lccc|ccc|ccc|c|}
					\hline
					& \multicolumn{3}{c|}{Bias settings}
					& \multicolumn{3}{c|}{at $t_\mathrm{int} = 5\,\upmu\mathrm{s}$}
					& \multicolumn{3}{c|}{at $t_\mathrm{int} = 9\,\upmu\mathrm{s}$}
					& \\
					\cline{2-10}
					Device
					& $V_\mathrm{GS,on}$
					& $\lvert V_\mathrm{DS}\rvert$
					& $\lvert I_\mathrm{DS}\rvert$
					& $\mathrm{ENC_{tot}}$
					& $\mathrm{ENC_{dep}}^{*}$
					& $\sigma_\mathrm{out}^{\dagger}$
					& $\mathrm{ENC_{tot}}$
					& $\mathrm{ENC_{dep}}^{*}$
					& $\sigma_\mathrm{out}^{\dagger}$
					& $g_q^{\ddagger}$ \\
					& [V]
					& [V]
					& [$\upmu\mathrm{A}$]
					& [$\mathrm{e}^-$]
					& [$\mathrm{e}^-$]
					& [pA]
					& [$\mathrm{e}^-$]
					& [$\mathrm{e}^-$]
					& [pA]
					& [$\mathrm{pA/e^-}$] \\
					\hline
					\StdL{}
					& $-2.40$ & $4.84$ & $134$
					& $1.56$ & $1.39$ & $841$
					& $1.43$ & $1.31$ & $769$
					& $539$ \\
					\StdR{}
					& $-2.75$ & $5.01$ & $102$
					& $1.68$ & $1.54$ & $932$
					& $1.51$ & $1.41$ & $840$
					& $555$ \\
					\BC{}
					& $1.19$ & $4.38$ & $51$
					& $1.09$ & $0.90$ & $687$
					& $0.96$ & $0.83$ & $606$
					& $632$ \\
					\SGQ{}
					& $-1.75$ & $2.81$ & $272$
					& $2.03$ & $1.98$ & $1695$
					& --- & --- & ---
					& $835$ \\
					\hline
				\end{tabular}
			}
		}
		\usebox{\biasenctable}
		\vspace{5pt}
		\parbox{\wd\biasenctable}{\footnotesize
			\textsuperscript{*}$\mathrm{ENC_{dep}}$ marks an upper estimate. It was calculated by quadratic subtraction of the $\mathrm{ENC_{elec}} (I_\mathrm{SSS} = 25\,\upmu\mathrm{A})$ from $\mathrm{ENC_{tot}}$.\\
			\textsuperscript{$\dagger$}$\sigma_\mathrm{out}$ is the $\mathrm{Noise_{out}}$ referred to the sensor output. \quad 
			\textsuperscript{$\ddagger$}$g_q$ is constant with $t_\mathrm{int}$.}
	\end{table}	
	
	The matrix homogeneity was compared at the respective longest integration time per device (figure~\ref{fig:IntTimeSweep} (c)). The \BC{} device exhibits the narrowest per-pixel noise distribution across the matrix, whereas \StdL{} and \StdR{} are in close agreement in both mode and dispersion once inter-production variations are taken into account. The \SGQ{} device shows a pronounced tail towards higher $\mathrm{ENC_{tot}}$ values, which is attributable to particularly noisy pixels, randomly distributed across the matrix.
	
	\begin{figure}[htbp]
	\centering
	\includegraphics[width=1\textwidth,trim=10 0 20 0,clip]{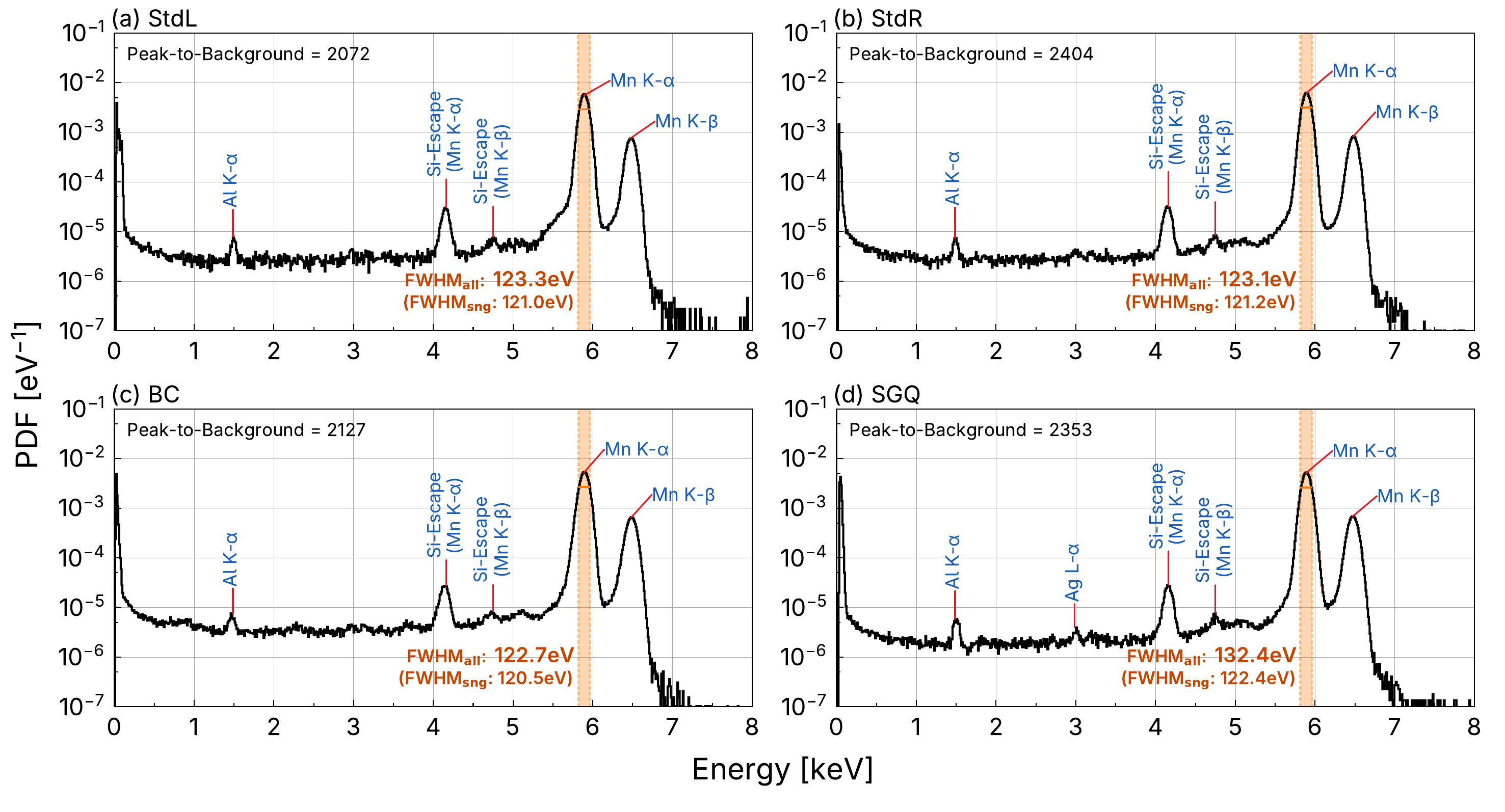}
	\caption{\label{fig:Spectrum} Normalised $^{55}\mathrm{Fe}$ energy spectra measured by the four devices at the bias settings of table~\ref{tab:bias_enc_summary}, their longest integration times, an exposure time of $2\,\mathrm{ms}$ and a sensor temperature of $158\,\mathrm{K}$.}
	\end{figure}
	At the respective longest integration times of all four devices, the $^{55}\mathrm{Fe}$ spectra show the characteristic manganese K$\alpha$ and K$\beta$ lines, as well as additional features such as the associated silicon escape peaks (figure~\ref{fig:Spectrum}). Across the devices the $\mathrm{FWHM_{all}}$ is comparable, except for the \SGQ{} device, whose higher $\mathrm{FWHM_{all}}$ was already observed in section~\ref{sec:noise_opt}. $\mathrm{FWHM_{sng}}$ is systematically lower than $\mathrm{FWHM_{all}}$, since split-frame events and the combined noise contributions from multiple pixels broaden the spectroscopic line. For all four devices, $\mathrm{FWHM_{sng}}$ approaches or falls within the resolution limit given by the Fano statistics ($120.1 \pm 0.7\,\mathrm{eV}$~\cite{loweMeasurementElectronHole2007b}) and the measured $\mathrm{ENC_{tot}}$, indicating excellent charge collection. At an exposure time of $2\,\mathrm{ms}$, the peak-to-background ratios are similar.\footnote{The ratio is defined as the Mn~K$\alpha$ peak divided by the background between $0.9$ and $1.1\,\mathrm{keV}$.} Increasing the exposure time to $50\,\mathrm{ms}$ reduces the fraction of misfits and split-frame events, improving the peak-to-background ratio fourfold. Longer exposure times led to pile-up at the given $^{55}\mathrm{Fe}$ source activity and distance to the sensor.

	\section{Discussion}
	\label{sec:discussion}
	\subsection{Novel DePFET technologies}
    For \BC{}, the results combine into a consistent picture of the intended reduction in coupling between the interface traps and the channel. This is supported by the lower $\sigma_\mathrm{out}$ at the longest integration time (section~\ref{sec:optimal}) and by the steep decrease in noise as the inversion layer at the interface is depleted (section~\ref{sec:noise_opt}). The permissible drain current before an inversion layer at the interface forms could, for future productions, be increased by enhancing the channel doping. The improved conductivity allows for a higher $g_\mathrm{m}$ and thus a higher $g_q$ (eq.~\eqref{eq:gq_gm}), while maintaining buried-channel operation. This is at the cost of shifting the onset of the impurity-scattering-related decrease in $g_q$ (section~\ref{sec:temperature}) to a higher temperature, and operation at higher current also entails higher power consumption. A desired operating range therefore depends more strongly on the channel implantation~\cite{huDesignTradeoffsSurface1985b, wilcoxLowtemperatureCharacterizationBuriedchannel1989}.

	For \SGQ{}, the main design predictions are largely met (sections~\ref{sec:noise_opt} and~\ref{sec:gain}), but the excess noise contribution outweighs the higher charge gain.\footnote{Targeted measurements on a second \SGQ{} device (not shown) confirm these results.} Its origin, however, cannot be determined conclusively from the conducted measurements. One candidate is additional traps introduced by the strip implant at the short source-side section, which may also explain the particularly noisy pixels. Another candidate is the position- and bias-dependent contribution of local noise sources within the laterally non-uniform channel (section~\ref{sec:designs})~\cite{mavredakisMeasurementCompactModeling2013b, kumarInvestigatingLateralTrap2025}. Which of these mechanisms is present or dominant determines the route to improving the super-$g_q$ technology, and assessing them requires detailed measurements of the noise spectral density accompanied by device simulations.
	
	\subsection{Limitations and common observations}
	All results are bounded by the experimental setup in several respects. The dynamic range of the Veritas(2.1) sets the longest accessible integration time and restricts DePFETs with increased internal amplification in particular. A readout ASIC with greater dynamic range or lower pre-gain settings, such as the latest Veritas(2.3), would extend the accessible integration time and allow measurements in which the thermal noise contribution is further suppressed~\cite{dakshinamurthyVERITAS231Optimisation2026b}. Furthermore, the noise of the readout chain itself, which is similar across all detector assemblies, becomes significant for devices with comparatively low sensor noise or high drain current, especially at short integration times. A thorough assessment of the electronics noise at the respective subtraction currents would allow a more accurate correction of the total noise and a comparison of the DePFET sensor noise across the different optimal bias settings and drain currents. Lastly, at the Mn~K$\alpha$ line, the measured energy resolution of all devices is essentially Fano-limited so that differences in ENC are barely reflected in the FWHM. The benefit of a low noise floor is therefore more evident at lower photon energies and for multi-pixel events, whose fraction increases towards smaller pixel pitches.
	
	Beyond these limitations, the comparison of the novel devices with \StdR{} is subject to not yet quantified variations between wafers and within individual wafers of the R\&D production. These variations remain to be assessed through measurements on further devices and dedicated test structures. The two standard devices, however, behave qualitatively alike, and the \StdL{} results agree with previous measurements~\cite{treberspurgCharacterization2562562020, bonholzerDrainCurrentCharacteristics2022, sommerCharacterisationNewAthenaWFIs2026b}.
	
	Common to all four devices is the loss in coupling efficiency at higher drain currents (section~\ref{sec:gain}). The first-order model of the charge gain (eq.~\eqref{eq:gq_gm}) therefore has to be extended such that $f$ is expressed as a function of the operating point, which is the subject of an ongoing study on the two-dimensional field effects caused by the internal-gate charge. Furthermore, studies are being conducted on the noise components and their temperature dependence, in particular to quantify the reduction of the $1/f$ contribution in \BC{} and to confirm that the noise increase below about $155\,\mathrm{K}$ originates from generation-recombination processes.
	
	\section{Summary and conclusions}
	\label{sec:summary}
	Within the course of this work, two novel DePFETs, the buried-channel device and the super-$g_q$ device, were characterised and compared with two standard DePFETs. Their noise behaviour and spectral properties were evaluated. The low-noise operating regions were determined, and the optimal bias conditions were identified. Furthermore, measurements were performed across a wide temperature range, and measurements at various integration times were conducted, including the assessment of the electronics noise contribution.
	
	Across all variants, a minimum noise level was observed at a sensor temperature of approximately $155\,\mathrm{K}$, where the single-pixel energy resolution remained limited by Fano statistics. Around this temperature, the buried-channel device achieved an ENC of $0.96\,\mathrm{e}^-$ at a readout time of $19.2\,\upmu\mathrm{s}$ per row, while requiring less than half the power consumption of the standard devices. Thus, to our knowledge for the first time, sub-electron noise has been reached in a single readout over a complete $64\times64$ DePFET matrix, even though a significant portion of the total noise is attributable to the readout electronics. Previously, $0.95\,\mathrm{e}^-$ had only been achieved with part of a matrix operated in window mode and at reduced dynamic range~\cite{treberspurgAchievableNoisePerformance2018a}. The present result was achieved in buried-channel operation and stems primarily from the reduced intrinsic noise and, to some extent, from the increased charge gain. However, the low-noise region is more limited than in the standard devices, as low noise requires the channel to remain fully buried. The super-$g_q$ device delivers the highest charge gain of all presented devices, approximately $1100\,\mathrm{pA/e^-}$, exceeding that of the reference device by a factor of 2.6 under matched bias conditions. Due to an excess noise contribution, however, its minimum measured ENC is higher than that of the standard devices, and the balance between noise and charge gain results in an optimal operating point in the transition regime, where the charge gain is reduced to $835\,\mathrm{pA/e^-}$.
	
	Overall, the results show that more fundamental changes to the intrinsic design, beyond conventional parameter optimisation, open up new degrees of freedom for tailoring the DePFET for future applications. 
	
	\appendix
	\section{Onset of impact ionisation}
	\label{sec:impact_ionisation}
\begin{figure}[htbp]
		\centering 
		\includegraphics[width=0.8\textwidth,trim=0 0 0 0,clip]{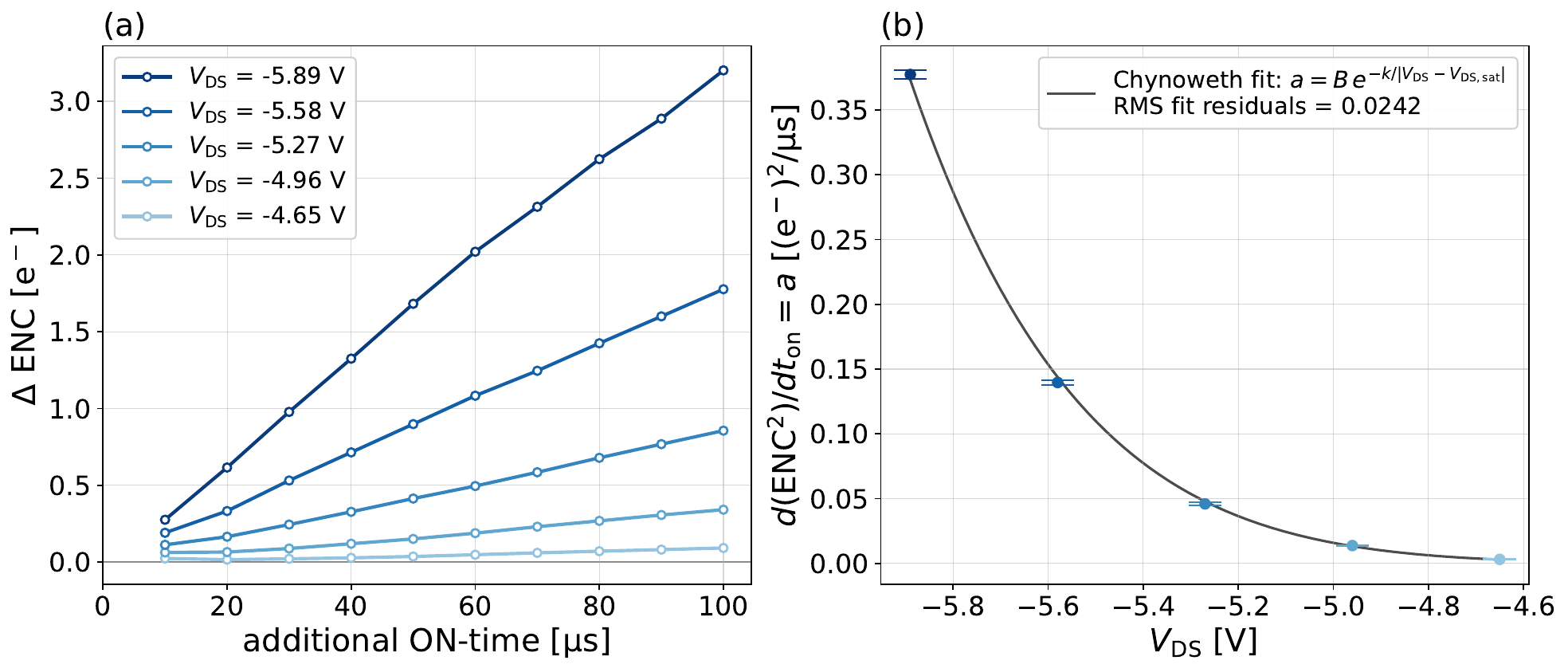}
		\caption{\label{fig:ImpIon} Impact-ionisation onset sweep of the \BC{} device. (a) Increase of ENC, relative to the fitted onset value at $t_\mathrm{on} = 0$, as a function of additional ON-time, for five $V_\mathrm{DS}$. (b) Slope $a$, obtained from linear fits to $\mathrm{ENC}^2(t_\mathrm{on})$, as a function of $V_\mathrm{DS}$, with a Chynoweth-law fit overlaid. Error bars represent the $\pm 1$ standard error of $a$ from the regressions.}
	\end{figure}

	To test whether the rapid increase of noise at high $\lvert V_\mathrm{DS} \rvert$ marks the onset of impact ionisation, a dedicated measurement was conceived using the \BC{} device as a representative example. For five operating points, an additional ON-time, varied from $t_\mathrm{on} = 10\,\upmu\mathrm{s}$ to $100\,\upmu\mathrm{s}$ in steps of $10\,\upmu\mathrm{s}$, was inserted before the readout process of each row. The increase in processing time per row was counterbalanced by a dwell time, keeping the frame cycle time and thereby the bulk leakage noise contribution to the measured total noise constant. As in the bias optimisation measurement (section~\ref{sec:bias2}), the drain current was constant at $\lvert I_\mathrm{DS}\rvert \approx 52\,\upmu\mathrm{A}$ and for a given $V_\mathrm{GS,on}$ a $V_\mathrm{DS}$ follows. The increase of ENC with ON-time for all five operating points is shown in figure~\ref{fig:ImpIon} (a). For a DePFET, the impact-ionisation generation rate is
	\begin{equation}
		G_\mathrm{II}
		=
		\frac{1}{q}\alpha_p(E)\,j_p,
		\label{eq:gii_holes}
	\end{equation}
	where $\alpha_p(E)$ is the hole ionisation coefficient and $j_p$ is the hole current density. The total impact-ionisation generation rate $R_\mathrm{II}$ is obtained by integrating over the high-field volume,
	\begin{equation}
		R_\mathrm{II}
		=
		\int_{\Omega_\mathrm{HF}} G_\mathrm{II}\,\mathrm{d}V
		=
		\frac{\lvert I_\mathrm{DS}\rvert}{q}
		\int_\mathrm{HF} \alpha_p(E(y))\,\mathrm{d}y.
		\label{eq:rii}
	\end{equation}
	Since $I_\mathrm{DS}$ is constant, the hole flux through the high-field region is constant. The variation of $R_\mathrm{II}$ across the operating points therefore arises from the electric-field $E$ dependence of $\alpha_p$, and thus from $V_\mathrm{DS}$. Electrons generated by impact ionisation drift to their local potential minimum, which for a DePFET is the internal gate ($\eta = 1$). Assuming further no avalanche multiplication at the onset of impact ionisation ($F_\mathrm{ex} = 1$), each event contributes one electron as signal charge, which follows Poisson counting statistics. The additional ON-time dependent noise contribution therefore measures the variance of these events and, since $\mathrm{Var}(N_\mathrm{II}) = \left\langle N_\mathrm{II}\right\rangle$, also the mean number of events,
	\begin{equation}
		\mathrm{ENC}^2(t_\mathrm{on})
		-
		\mathrm{ENC_0}^2
		=
		\eta F_\mathrm{ex}
		\left\langle N_\mathrm{II}\right\rangle
		=
		\eta F_\mathrm{ex}
		R_\mathrm{II}\,t_\mathrm{on}
		=
		R_\mathrm{II}\,t_\mathrm{on},
		\qquad
		\substack{
			\eta\,\coloneqq\,1\ \text{(collection efficiency)} \\
			F_\mathrm{ex}\,\coloneqq\,1\ \text{(excess-noise factor)}
		}
		\label{eq:enc_ton_ii}
	\end{equation}
	where $\mathrm{ENC_0}$ is the ON-time independent contribution. Hence, the slope $a$ of $\mathrm{ENC}^2$ as a function of	$t_\mathrm{on}$ measures the impact-ionisation generation rate,
	\begin{equation}
		a
		=
		\frac{\mathrm{d}(\mathrm{ENC}^2)}
		{\mathrm{d}t_\mathrm{on}}
		=
		R_\mathrm{II}.
		\label{eq:enc_slope_ii}
	\end{equation}
	Figure~\ref{fig:ImpIon} (b) shows the resulting slope for each operating point. To obtain the expected $V_\mathrm{DS}$ dependence, the high-field region is approximated by an effective length $d_\mathrm{eff}$ across which the remaining potential drop occurs, such that $E \approx (\lvert V_\mathrm{DS} - V_\mathrm{DS,sat}\rvert)/d_\mathrm{eff}$. Here, $V_\mathrm{DS,sat} = -1.2\,\mathrm{V}$ was determined from the transition to the linear regime along the corresponding constant current line in the $V_\mathrm{GS,on}$--$V_\mathrm{DS}$ plane (section~\ref{sec:bias2}). With Chynoweth's empirical law,
	\begin{equation}
		\alpha_p(E)
		=
		\alpha_{\infty,p}
		\exp\left(-\frac{b_p}{E}\right),
		\label{eq:chynoweth_holes}
	\end{equation}
	equations~\eqref{eq:rii} and~\eqref{eq:enc_slope_ii} yield
	\begin{equation}
		a(V_\mathrm{DS}) = B \exp\left( -\frac{k}{\lvert V_\mathrm{DS} - V_\mathrm{DS,sat}\rvert} \right),
		\label{eq:chynoweth_rate}
	\end{equation}
	with $k = b_p d_\mathrm{eff}$ and $B$ a prefactor~\cite{chynowethIonizationRatesElectrons1958}. Using the hole impact-ionisation parameter $b_p = 2.35\times10^{6}\,\mathrm{V\,cm^{-1}}$ at $220\,\mathrm{K}$ from the expanded Massey model of Currás-Rivera and Moll~\cite{curras-riveraTemperatureDependenceAvalanche2026a}, the measured constant $k = 63 \pm 1\,\mathrm{V}$ corresponds to an effective high-field length $d_\mathrm{eff} = 0.27\,\upmu\mathrm{m}$. A physical high-field length is presumably shorter than $d_\mathrm{eff}$ due to the homogeneous high-field assumption. It would, however, require dedicated measurements and simulations to determine the lateral field profile and the remaining potential drop precisely, including variation of $V_\mathrm{DS,sat}$ caused by the $V_\mathrm{GS,on}$ adjustment required to keep $I_\mathrm{DS}$ constant. Nonetheless, the linear increase of $\mathrm{ENC}^2$ with the ON-time, as expected for a constant generation rate, together with the exponential field dependence through $V_\mathrm{DS}$ and a reasonable $d_\mathrm{eff}$ relative to the channel length, supports impact ionisation as the origin of the additional noise at high $\lvert V_\mathrm{DS} \rvert$.
	
	\section{Veritas gain exploration}
	\label{app:GainExplo}
\begin{figure}[htbp]
		\centering 
		\includegraphics[width=1\textwidth]{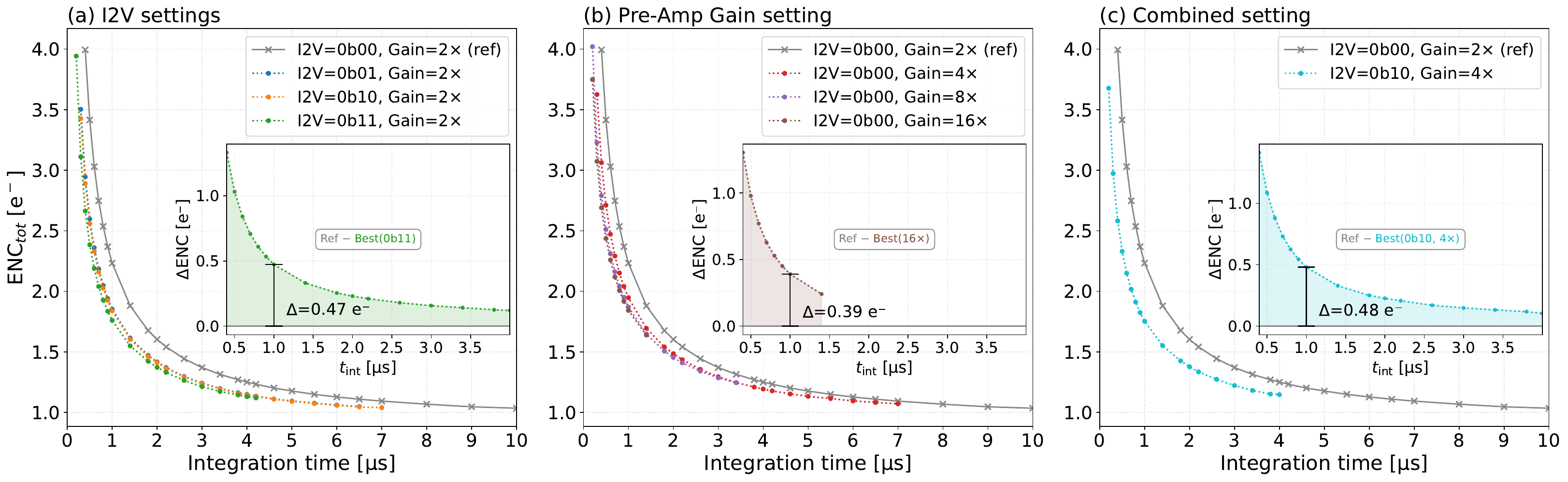}
		\caption{\label{fig:VeritasSettings} Impact of various Veritas gain settings on noise, using the \BC{} device as an example, at the bias conditions of section~\ref{sec:optimal}. Total noise $\mathrm{ENC_{tot}}$ as a function of integration time $t_\mathrm{int}$ for increased I2V feedback resistance (a), increased preamplifier gain (b), and the combined setting (c), each compared with the reference configuration (i.e. lowest available Veritas gain settings). The insets show the noise reduction $\Delta\mathrm{ENC}$ of the respective best setting relative to the reference.}
	\end{figure}

	Due to the significant contribution of $\mathrm{ENC_{elec}}$ at short integration times $t_\mathrm{int}$, particularly relevant for fast readout requirements, the influence of various Veritas gain settings was evaluated using the \BC{} device under the same bias conditions as described in section~\ref{sec:optimal}. With increasing Veritas gain, the dynamic range of the integration stage is reached earlier, and the accessible integration time, for a given signal current, decreases accordingly. At an integration time of $t_\mathrm{int} = 1\,\upmu\mathrm{s}$, an $\mathrm{ENC_{tot}}$ of $2.25\,\mathrm{e}^-$ is obtained in the reference configuration (i.e. lowest Veritas gain settings). Maximising the I2V feedback resistance ($\mathrm{I2V} = 0\mathrm{b}11$, table~\ref{tab:I2V}) lowers the noise by $\Delta\mathrm{ENC} = 0.47\,\mathrm{e}^-$ (figure~\ref{fig:VeritasSettings} (a)), and increasing the preamplifier gain to $16\times$ by $\Delta\mathrm{ENC} = 0.39\,\mathrm{e}^-$ (figure~\ref{fig:VeritasSettings} (b)). In both cases, progressively higher gains result in diminishing improvements. A combined configuration ($\mathrm{I2V} = 0\mathrm{b}10$, $\mathrm{gain} = 4\times$) performs comparably to the maximised I2V setting (figure~\ref{fig:VeritasSettings} (c)). The benefit of a higher Veritas gain is most pronounced at the shortest integration times, reaching $\Delta\mathrm{ENC} = 1.40\,\mathrm{e}^-$ at $t_\mathrm{int} = 0.4\,\upmu\mathrm{s}$.
	\begin{table}[htbp]
		\centering
		\caption{\label{tab:I2V} Veritas I2V stage settings with the corresponding feedback resistance $R$ and capacitance $C_\mathrm{c}$~\cite{herrmannVERITAS22Low2018a}.}
		\smallskip
		\begin{tabular}{|c|c|c|}
			\hline
			I2V setting & $R$ [k$\Omega$] & $C_\mathrm{c}$ [fF] \\
			\hline
			$0\mathrm{b}00$ & 50  & 190 \\
			$0\mathrm{b}01$ & 100 & 100 \\
			$0\mathrm{b}10$ & 100 & 190 \\
			$0\mathrm{b}11$ & 150 & 100 \\
			\hline
		\end{tabular}
	\end{table}
	
	\section{Complete DePFET transistor bias sweep results}
	\label{app:CCmatrix}
	The following figures~\ref{fig:matrixAthena}--\ref{fig:matrixSuper} present the complete results of the DePFET transistor bias sweep of section~\ref{sec:bias2} for the four DUTs respectively. The scatter matrix of each figure shows the pairwise relations between the voltages and the measured quantities, whereas the inset maps $\mathrm{Noise_{out}}$ in the $V_\mathrm{GS,on}$--$V_\mathrm{DS}$ plane, analogous to the ENC map of figure~\ref{fig:overviewGAIds}. The colour scale of the inset as well as colour denoting constant $I_\mathrm{DS}$ in the matrix scatter plots are common among the figures. The energy calibration gain ($G_\mathrm{E}$) relates to the charge gain ($g_q$) by a factor of $w(T)/A = 275\,(\mathrm{pA/e^-})/(\mathrm{ADU/eV})$.
	
	\begin{figure}[htbp]
		\centering 
		\includegraphics[width=1\textwidth,trim=0 0 0 0,clip]{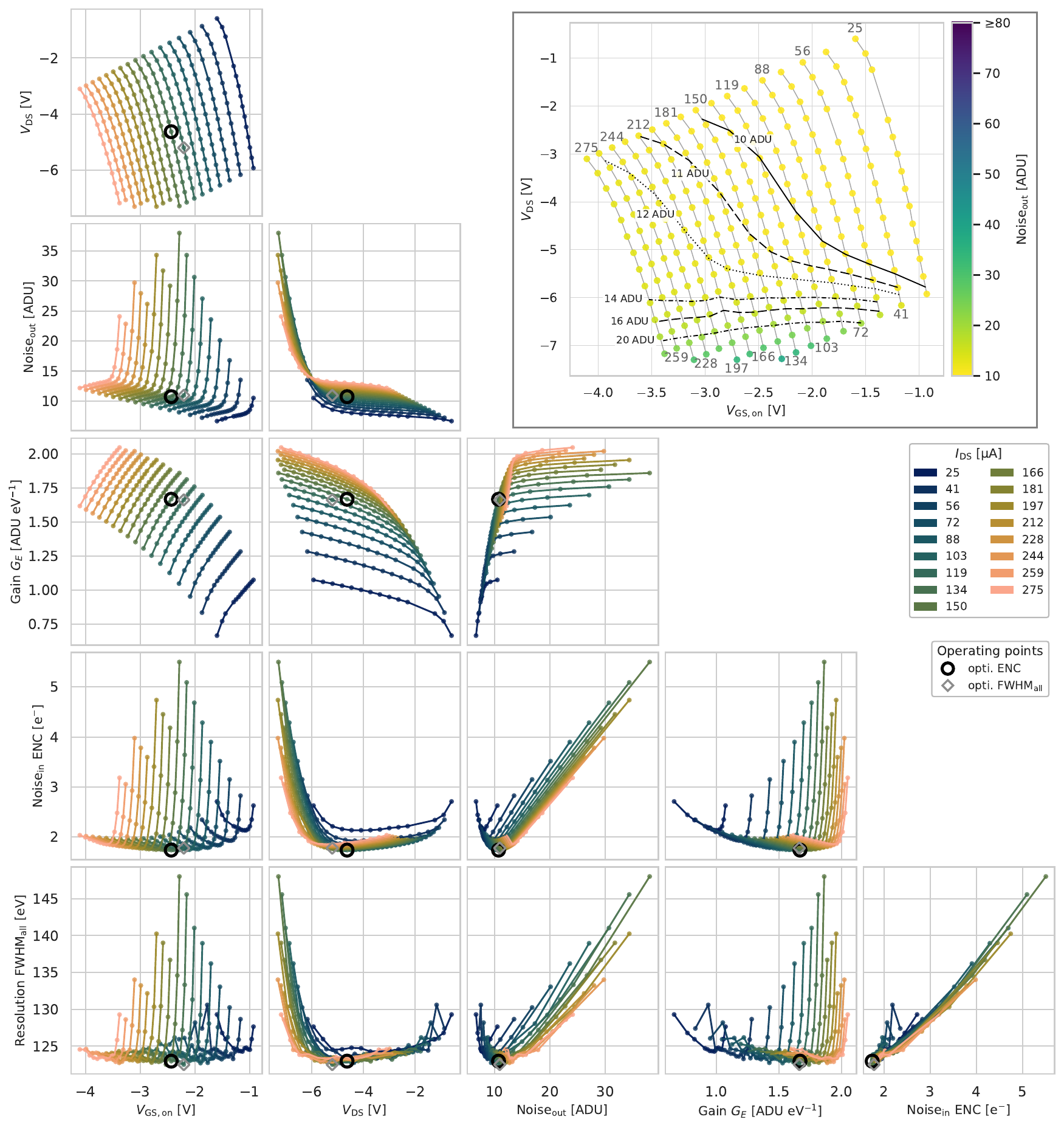}
		\caption{\label{fig:matrixAthena} All valid measurement points of the DePFET transistor bias sweep of the \StdL{} device (section~\ref{sec:bias2}). Scatter matrix: pairwise relations between dependent ($\mathrm{Noise_{out}}$, $G_\mathrm{E}$, ENC, $\mathrm{FWHM_{all}}$, $V_\mathrm{DS}$) and independent variables ($V_\mathrm{GS,on}$, $I_\mathrm{DS}$). Each colour denotes one constant $I_\mathrm{DS}$ contour (legend, in $\upmu\mathrm{A}$). The circle marks the ENC optimum and the diamond the $\mathrm{FWHM_{all}}$ optimum of the swept bias space. Inset (top right): $\mathrm{Noise_{out}}$ in the $V_\mathrm{GS,on}$--$V_\mathrm{DS}$ plane, with grey lines of constant $I_\mathrm{DS}$ (labelled in $\upmu\mathrm{A}$) and black iso-lines of $\mathrm{Noise_{out}}$ (labelled in ADU).}
	\end{figure}
	\begin{figure}[htbp]
		\centering 
		\includegraphics[width=1\textwidth]{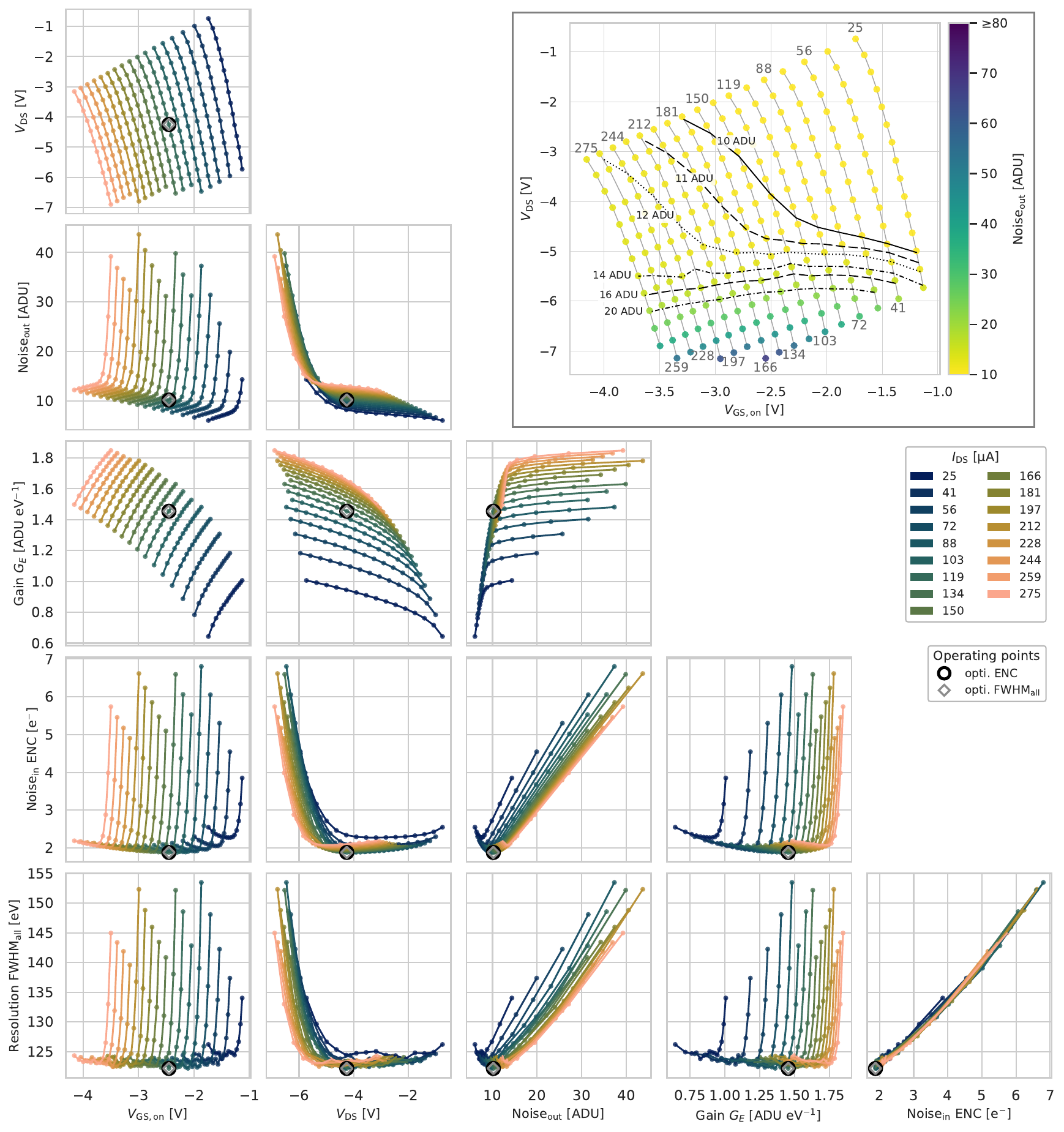}
		\caption{\label{fig:matrixRef} As figure~\ref{fig:matrixAthena}, for the \StdR{} device. The ENC and $\mathrm{FWHM_{all}}$ optima coincide.}
	\end{figure}
	\begin{figure}[htbp]
		\centering 
		\includegraphics[width=1\textwidth]{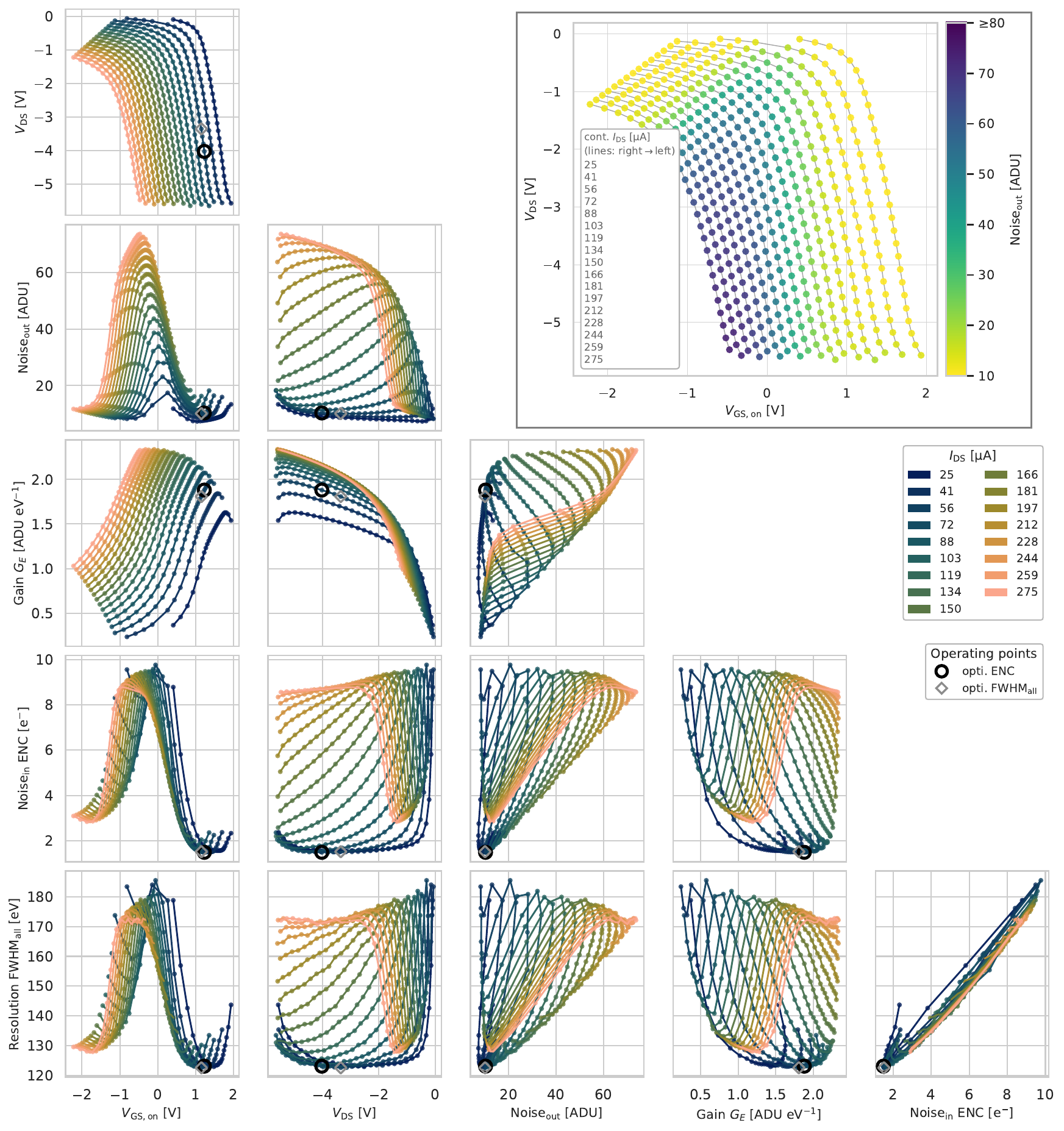}
		\caption{\label{fig:matrixOpti} As figure~\ref{fig:matrixAthena}, for the \BC{} device. No $\mathrm{Noise_{out}}$ iso-lines are drawn, since $\mathrm{Noise_{out}}$ does not vary monotonically with $V_\mathrm{GS,on}$ at constant $I_\mathrm{DS}$.}
	\end{figure}
	\begin{figure}[htbp]
		\centering 
		\includegraphics[width=1\textwidth]{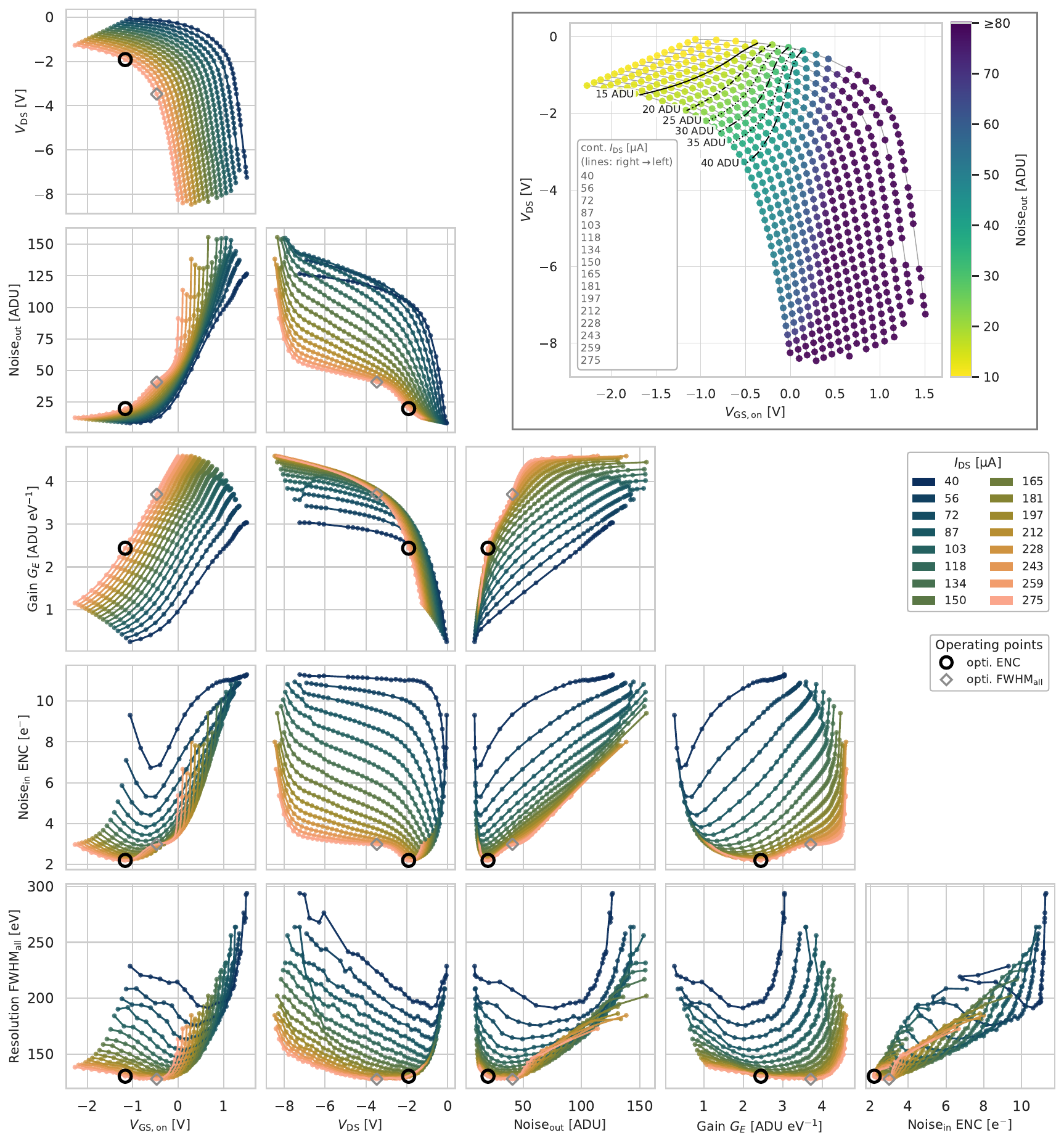}
		\caption{\label{fig:matrixSuper} As figure~\ref{fig:matrixAthena}, for the \SGQ{} device.}
	\end{figure}
	
	\clearpage
	\acknowledgments
	The authors thank the technical staff of HLL, in particular Gerhard Schaller for processing and for overseeing the PXD13 production, Danilo Mießner for assembly, and Martin Hensel for system integration. S. Stadler also thanks Prof. Alfred Kersch (Munich University of Applied Sciences) for his guidance, valuable insights and support throughout this study. 
		
	\paragraph{Data availability.} The data supporting the findings of this study are available from the corresponding author upon reasonable request.
	
	\paragraph{AI disclosure.} ChatGPT (OpenAI) and Qwen were used to check the manuscript for consistency and language and to support drafting the plots in Python. These tools did not generate or modify any measurement data. The authors reviewed, validated and critically assessed all AI-suggestions and take full responsibility for the content of the manuscript.
	
	\bibliographystyle{JHEP}
	\bibliography{JINST}

\end{document}